\documentclass[
 reprint,
 aps,
 prl,
 amsmath,amssymb,
 aps,
]{revtex4-2}

\usepackage{graphicx}
\usepackage{dcolumn}
\usepackage{bm}
\usepackage[font=small,labelfont=bf, justification=justified,format=plain]{caption}
\usepackage{adjustbox}
\usepackage{siunitx}

\usepackage{comment}

\begin{document}

\preprint{APS/123-QED}

\title{A universal reconstruction method for X-ray scattering tensor tomography based on wavefront modulation}
\author{Ginevra Lautizi\textsuperscript{1,2}}
\email{ginevra.lautizi@phd.units.it}
\author{Alain Studer\textsuperscript{3}}%
\author{Marie-Christine Zdora\textsuperscript{4,5}}
\author{Fabio De Marco\textsuperscript{1,2}}
\author{Jisoo Kim\textsuperscript{6}}
\author{Vittorio  Di Trapani\textsuperscript{1,2}}
\author{Federica Marone\textsuperscript{4}}
\author{Pierre Thibault\textsuperscript{1,2}}
\author{Marco Stampanoni\textsuperscript{4,5}}
\email{marco.stampanoni@psi.ch}
\affiliation{\textsuperscript{1} Department of Physics, University of Trieste, Trieste, Italy }
\affiliation{\textsuperscript{2} Elettra-Sincrotrone Trieste, Basovizza, Italy}
\affiliation{\textsuperscript{3} Data Processing Development and Consulting Group, Paul Scherrer Institut, Villigen 5232, Switzerland}
\affiliation{\textsuperscript{4} Photon Science Division, Paul Scherrer Institut, Villigen 5232, Switzerland}
\affiliation{\textsuperscript{5} Institute for Biomedical Engineering, ETH Zürich, Zürich 8092, Switzerland}
\affiliation{\textsuperscript{6}Advanced Instrumentation Institute, Korea Research Institute of Standards and Science, Korea}


\begin{abstract}
We present a versatile method for full-field, X-ray scattering tensor tomography that is based on energy conservation and is applicable to data obtained using different wavefront modulators. Using this algorithm, we pave the way for speckle-based tensor tomography. The proposed model relies on a mathematical approach that allows tuning spatial resolution and signal sensitivity. We present the application of the algorithm to three different imaging modalities and demonstrate its potential for applications of X-ray directional dark-field imaging. 

\end{abstract}

\maketitle

\section{INTRODUCTION}
The micro- and nanostructures inside a macroscopic object play a fundamental role in its characteristics and properties. For instance, the mechanical properties of bones are closely linked to the alignment of collagen fibers at a local level \cite{Fratzl2008}. Likewise, the mechanical characteristics of fiber-reinforced polymers and composites are significantly influenced by the local orientation of synthetic fibers \cite{Pipes1982}.

Small-angle X-ray scattering (SAXS) tensor tomography can provide information about the alignment of microstructures of non-crystalline samples on a sub-micron length scale, often not resolvable with conventional approaches such as micro-CT \cite{Schaff2015, Liebi2015, Liebi2018, Gao2019, Skjonsfjell2016}.
However, scanning SAXS requires long acquisition times due to drastic reduction in flux to create a narrow beam that has to be scanned over the sample at different sample angular poses. The long acquisition times of scanning SAXS, combined with long analysis times, limit its applications, in particular for large samples.
For these reasons, the established approaches require experimental times in the order of several hours, even for small sample volumes.

Full-field X-ray imaging methods capable of delivering directional information about sub-structures in a sample, such as speckle-based imaging (SBI) and grating interferometry (GI), have been actively investigated in recent years. 
Both GI and SBI techniques make use of a wavefront-marking element, such as gratings and sandpaper respectively, to generate intensity fluctuations in the recorded detector image \cite{Momose2003, Weitkamp2005}. The local distortion of the pattern, introduced by the sample, is then analyzed to retrieve attenuation, phase contrast, and scattering (dark-field) information \cite{Pfeiffer2008}.
SBI has become more and more common in recent years. It is based on a similar principle as GI and it has shown similar scattering sensitivities, without the need of tailored phase optical elements \cite{Morgan2012, Berujon2012, Zanette2015, Zdora2018, Zhou2018, Zdora2021}.
Compared to SAXS imaging, these full-field techniques allow for larger fields of view (FOVs) to be investigated in shorter measurement times. Moreover, access to the dark-field information is not directly limited by the beam properties, such as size, monochromaticity and transverse coherence \cite{Yashiro2010, Lynch2011, Strobl2014}.
Given that the directional dark-field signal provides information on the sub-structures' orientations in a sample, it becomes possible to detect these structures, even when their dimensions are smaller than the image pixel resolution, enabling a reliable reconstruction of the prevalent orientations of microstructural features \cite{Malecki2014, Vogel2015, Jensen2010, Kim2022b, Smith2022}.

Recently, 2D omnidirectional X-ray scattering sensitivity in a single shot has been demonstrated using circular gratings, paving the way for time-resolved studies \cite{Kim2020, Kim2021, Kim2022}. This study combines an improved acquisition speed based on circular diffractive optics with an optimized data acquisition protocol. A tomographic reconstruction method based on a filtered back-projection with algebraic filters has also been introduced.
However, in this approach the extraction of dark-field signal relies on a tailored circular phase-grating array that requires specialized facilities and knowledge to be fabricated. Moreover, the use of such a grating array does not allow for easy tuning of the X-ray energy and spatial resolution of the scattering images. A general reconstruction method is therefore needed, where the dark-field signal extraction is not strictly dependent on the type of wavefront marker used.

\section{METHOD}
In this manuscript, we suggest a general reconstruction method that can be applied both to GI and SBI X-ray scattering approaches to obtain tensor tomography volumes. Unlike existing algorithms \cite{Felsner2019, Graetz2021}, our method is based on the mathematical rotation of the scattering tensor and can therefore be applied to different techniques. Moreover, we reformulated the reconstruction problem directly and intuitively, leading to the reconstruction of the full tensor field, not only the scattering signal along predefined directions.

\subsection{Omnidirectional dark-field signal extraction}

Our method is based on energy conservation: it assumes that the intensity of the photon distribution not absorbed by the sample is conserved by the scattering process. The model considers a wave field intensity distribution generated by an ideal point source.

The scattering can be modeled, on a first approximation, as a Gaussian probability distribution. The process underlying the X-ray dark-field image modality is ultra-small angle X-ray scatter (USAXS), which is a coherent process. In SAXS/USAXS theory, the scattered fields from all regions a given microstructure are integrated, which leads to a certain distribution of scattered fields as a function of the scattering vector $\bm{q}$, or equivalently, the scattering angle $\theta$. The squared magnitude of the field gives the scattered intensity as a function of the scattering angle. The relationship with $\bm{q}$ or $\theta$ is not generally a Gaussian, but often resembles one. In SAXS, if sufficiently small $\bm{q}$ values are probed, approximating $I(\bm{q})$ with a Gaussian, the Guinier approximation, is often used to retrieve the radius of gyration, a measure for the particle size. While the Guinier approximation does not accurately model the behavior at large scattering vectors $\bm{q}$, using the approximation is almost always appropriate in imaging, because the relative scattering intensities at high scattering vectors are very small ($< 10^{-3}$). In an imaging geometry, scattered and unscattered beams overlap spatially, and thus high-$\bm{q}$ scattering is effectively invisible.

The covariance matrix of the Gaussian distribution can be interpreted as the scattering tensor of the sample. The physical model used can be derived from the Photon Diffusion equation \cite{Aronson1999} combined with the Fokker-Planck model for X-ray imaging \cite{Paganin2019}.
Therefore, if we label the incident beam with $I_{0}$ and the transmitted beam leaving the voxel with $I_{s}$, for each unit cell we can write 
\begin{equation}
I_{s}(\bm{r_\perp}) = I_{0}(\bm{r_\perp})e^{-\mu(y)\Delta y}  * |\Sigma_y|^{-\frac{1}{2}}e^{-\frac{1}{2}\bm{r_\perp}^\intercal \Sigma_y^{-1} \bm{r_\perp}}\label{2d_real},
\end{equation}
where $\Sigma_y$ is the 2D scattering tensor (covariance matrix) of a thin slice at position $y$ in the beam, as the beam propagates along the $y$ direction (Fig.~\ref{fig:fig0}a). The 2D pixel coordinates in each unit cell are indicated as $\bm{r_\perp}$, where $\bm{r_\perp}$ represents the 2D coordinates transverse to the beam direction: $(x,z)$ in Fig.~\ref{fig:fig1}b.
The convolution operator is indicated as $*$, $\mu(y)$ is the absorption coefficient for each unit cell at a given position $y$, and $\Delta y$ is the thickness of the unit cell. The unit cell is given by the period of the pattern for the grating array and by an arbitrary window for a non-periodic wavefront modulator, such as sandpaper.
Performing the Fourier transform on Eq.~\ref{2d_real}, taking the absolute value, and the natural logarithm, it follows that
\begin{equation}
\frac{1}{2} \bm{k_\perp} \Sigma_y \bm{k_\perp} +\mu(y)\Delta y = - \ln \frac{ |\hat I_{s}(\bm{k_\perp})| }{|\hat I_{0}(\bm{k_\perp})|}.
\label{2d_fourier}
\end{equation}
The right hand side of Eq.~\ref{2d_fourier} is always non-negative, hence the left hand side must be a positive definite bilinear form in the variable $\bm{r_\perp}$. For each projection the 2D scattering tensor can be extracted in Fourier space, fitting each arbitrarily defined analysis window with a positive definite bilinear form, using a multislice version of Eq.~\ref{2d_fourier}.

Eq.~\ref{2d_real} can be generalized to a multislice version, considering one additional convolution per voxel:

\begin{equation}
I(\bm{r_\perp}) = I_{0}(\bm{r_\perp})e^{-\int_0^y \mu(y') dy'}  * 
e^{-\frac{1}{2}\bm{r_\perp}^\intercal \Sigma_0^{-1} \bm{r_\perp}}
\, * \, ... \, *
e^{-\frac{1}{2}\bm{r_\perp}^\intercal \Sigma_y^{-1} \bm{r_\perp}},
\label{2d_real_multislice}
\end{equation}
yielding the magnitude of the Fourier transform
\begin{equation}
|\hat I(\bm{k_\perp})| = 
|\hat I_{0}(\bm{k_\perp})|e^{-\int_0^y \mu(y') dy'}   
 e^{-\frac{1}{2} \bm{k_\perp} \int_0^y \Sigma(y') dy' \bm{k_\perp}}.
\label{2d_fourier_multislice}
\end{equation}

Since we cannot see scattering in beam direction, the full 3D $\Sigma$ needs to be corrected by a projection
operator $P_y, P_y \Sigma := P_B \Sigma P_B^T$ such that

\begin{equation}
\frac{1}{2}\bm{k_\perp}  \int_0^y P_y \Sigma(y') dy' \, \bm{k_\perp} + \int_0^y \mu(y') dy' = 
- \ln \frac{ |\hat I(\bm{k_\perp})| }{ |\hat I_{0}(\bm{k_\perp})| }.
\label{2d_fourier_multislice_2}
\end{equation}
The projection operator is derived in \cite{supplementary}.

\subsection{The mathematics of tensor tomography}
After having extracted the omnidirectional dark-field signal for each projection, we need to introduce a new mathematical model to reconstruct the tensor tomogram. 
Let $F$ be a symmetric, positive definite tensor field, $F: \mathbb{R}^{3} \rightarrow \mathbb{R}^{3 \times 3}$.
For any rotation $R_{angle}^{axis} \in SO(3)$, $D(R_{angle}^{axis})$ maps a tensor field $F$ to the rotated tensor field. For positions $\bm{r} = (x,y,z)$ and for our specific scan protocol, we can write
 
\begin{equation}
D(R_{\beta}^x)D(R_{\alpha}^z )[F(\bm{r})] =
(R_{\beta}^x R_{\alpha}^z) \cdot F\big( (R_{\beta}^x R_{\alpha}^z)^{-1} \bm{r} \big) 
\cdot (R_{\beta}^x R_{\alpha}^z)^{-1}.
\end{equation}

Having defined how a tensor field transforms, the definition of a tensor 
sinogram is straightforward. 
The 2-axes sinogram $S = A[F]$ is defined as 
\begin{equation}
A[F](P_B \bm{r}, \alpha, \beta) = \int_{\mathbb{R}} P_B D(R_{\beta}^x)D(R_{\alpha}^z )[F(x,y,z)] P_B dy ,\label{eq:tomoA}
\end{equation}
where $P_B$ is the projection operator along the beam propagation direction.

Reconstructing the tensor tomogram means solving the linear system $AX =B$, where $X$ denotes the tensor field to be reconstructed, $B$ is the measured tensor sinogram and $A$ the linear operator defined above. This linear system can be solved with conjugate gradient method (CG) or least squares (LS). Both the forward tensor operator $A$ and the adjoint $A^*$ are needed for the reconstruction. It can be proven (\cite{supplementary}) that $A^*$ equals

\begin{eqnarray}
&&A^*[S](\bm{r}) =
 \int_0^{2\pi} \int_0^{\pi/2}
 (R_{\beta}^x R_{\alpha}^z)^{-1} S(P_B R_{\beta}^x R_{\alpha}^z \bm{r}, \alpha, \beta)  R_{\beta}^x R_{\alpha}^z\nonumber\\
&&\times
\, \cos \beta d \beta d \alpha.\label{subeq:A*}
\end{eqnarray}

With the method we present, it is possible to retrieve the tensor tomogram also using a single tilt axis. However, having multiple tilt axes, shrinks the nullspace of the tensor tomography forward operator, enhancing the convergence of the reconstruction algorithm \cite{Kim2021}.

Once the scattering tensor field is reconstructed, it is then eigendecomposed. Knowing the eigenvalues of the scattering tensor for each voxel, the following quantities can be defined.
The mean of the three eigenvalues represents the mean scattering $\mathrm{MS} = (\lambda_1 + \lambda_2 + \lambda_3)/3$.
The fractional anisotropy ($\mathrm{FA}$) \cite{Basser1996}, which represents the scattering anisotropy, describes how well-aligned the fibers are in each voxel:
\begin{equation}
\mathrm{FA} = \frac{\sqrt{(\lambda_1-\lambda_2)^2+(\lambda_2-\lambda_3)^2+(\lambda_3-\lambda_1)^2}}{\sqrt{2(\lambda_1^2+\lambda_2^2+\lambda_3^2)}}.
\end{equation}
Areas characterized by a high volume fraction of well-aligned fibers will exhibit a high $\mathrm{FA}$.
Finally, the eigenvector with the lowest eigenvalue represents the preferential local fiber orientation, since the scattering is typically weakest along the fiber orientation. 
We will therefore show the above-mentioned quantities, as meaningful indicators accessible through tensor tomography for assessing the level of fiber scattering, alignment, and orientation.

\section{EXPERIMENTS AND RESULTS}
\begin{figure}[h!]
    \centering
    \includegraphics[width=\linewidth]{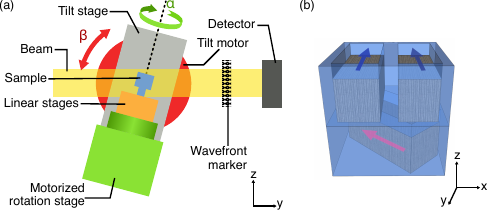}
    \caption{(a) Experimental setup outline viewed from the side and (b) a 3D model of the validation sample, where carbon fibers lie along two main orientations (arrows) inside a PMMA box.}
    \label{fig:fig0}
\end{figure}

We validated the method for use with full-field imaging methods utilizing three different wavefront markers: a circular phase-grating array, a fractal wavefront modulator \cite{Shi2022}, and a sandpaper diffuser.
The projection images with both the gratings and the fractal pattern were acquired at the TOMCAT beamline (Swiss Light Source, Paul Scherrer Institut) following the stair-wise acquisition protocol described in \cite{Kim2020, Kim2021}, using a monochromatic X-ray beam of $ \SI{17}{\kilo\eV}$. For each tilt angle $\beta$, we acquired $1000$ projections with a continuous rotation of the sample over $\alpha \in [\SI{0}{\degree},\SI{360}{\degree}$], while $\beta \in [\SI{0}{\degree},\SI{24}{\degree}$] with an angular step of $\SI{2}{\degree}$ (Fig.~\ref{fig:fig0}a).

We fabricated the test sample shown in Fig.~\ref{fig:fig0}b. The sample has a size of $ 4 \times 4 \times 4\ \SI{}{\mm^3}$, and is composed of two superimposed PMMA blocks with cavities, which contain bundled carbon fibers with a diameter of $\SI{12}{\micro\metre}$, oriented along the $y$-axis and at $\SI{45}{\degree}$.

\subsection{Circular gratings array}
The first scan was carried out using a $\pi$-shifting circular gratings array \cite{Kagias2016, Kagias2019} with a unit cell period $P = \SI{49.5}{\micro\metre}$ and a fine grating period $g = \SI{1.46}{\micro\metre}$. We used the CMOS-based GigaFRoST \cite{Mokso2017} detector coupled with a high-numerical-aperture tandem 1:1 microscope optic leading to an effective pixel size of $\SI{11}{\micro\metre}$. The detector was placed at a distance of $\SI{49.5}{\cm}$ downstream of the grating array, while the grating-sample distance was $\SI{46.3}{\cm}$. 
The exposure time for each projection was $\SI{12}{\milli\second}$ and the FOV was $460 \times 1008$ pixels.
The unit cell for the analysis, i.e. the image spatial resolution, corresponds to a square whose size is the diameter of a single circular grating: $9$ pixels or $\SI{99}{\micro\metre}$.

\begin{figure}[h!]
    \centering
    \includegraphics[width=\linewidth]{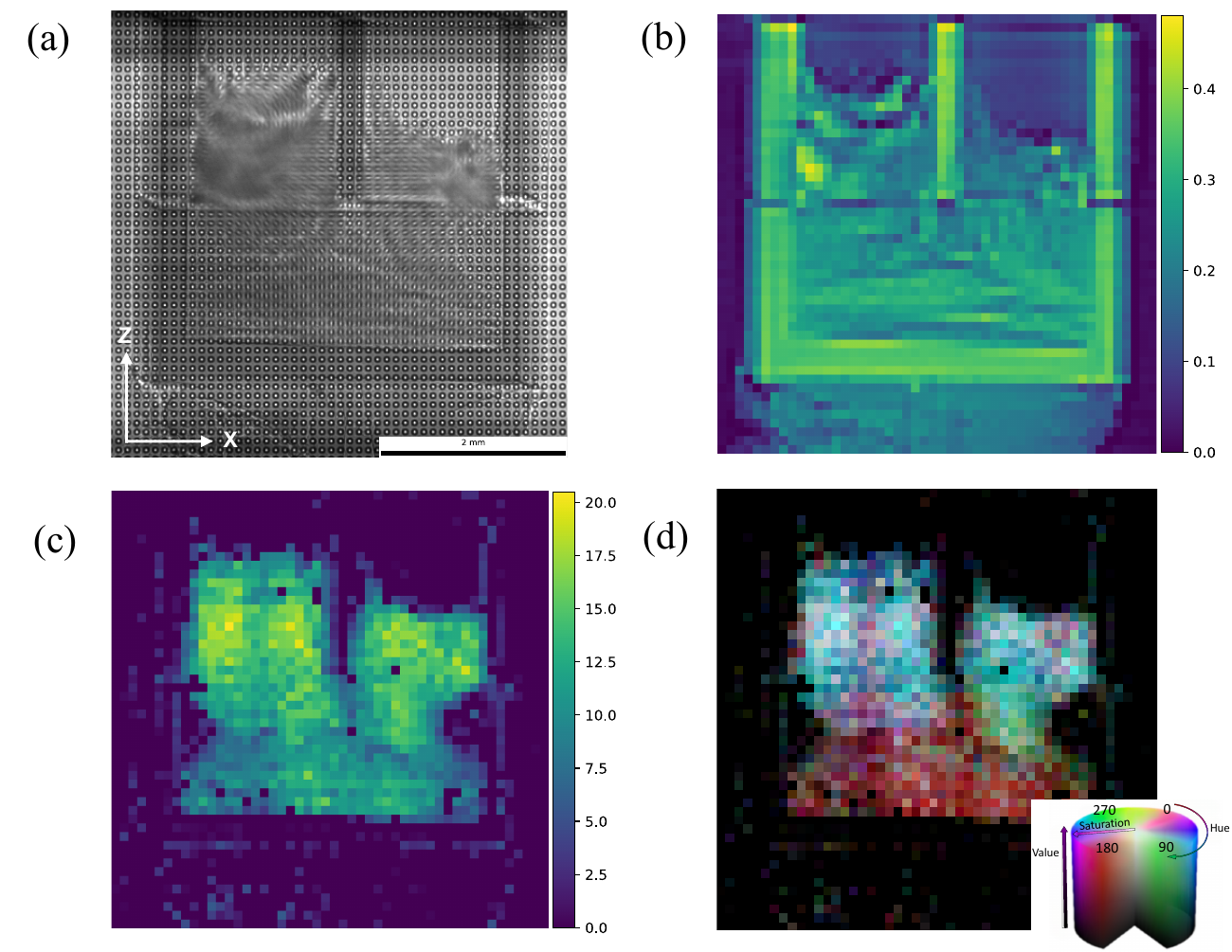}
    \caption{(a) Raw transmission image of the validation sample with a grating array at $\alpha = \SI{0}{\degree}$,  $\beta = \SI{0}{\degree}$. (b) Extracted absorption image, (c) mean scattering in arbitrary units and (d) main orientation signals. Each pixel corresponds to one unit cell.  $\SI{0}{\degree}$ (red) corresponds to the horizontal direction ($x$) in (a).}
    \label{fig:fig1}
\end{figure}

A raw transmission image of the validation sample, obtained with a single shot, is shown in Fig.~\ref{fig:fig1}a. In the image, the circular fringes of the gratings are visible, and their blurring due to strong scattering from the sample can be observed. The extracted absorption image is shown in Fig.~\ref{fig:fig1}b, while the mean scattering (Fig.~\ref{fig:fig1}c) is calculated starting from Eq.~\ref{2d_fourier}, as the mean of the eigenvalues of the scattering tensor. The main orientation (Fig.~\ref{fig:fig1}d) is the HSV representation of the 2D eigenvectors with the shortest lengths, where the hue (color shade) is the fiber orientation projected onto the detector plane, the saturation is the fractional anisotropy, and the value (brightness) is the mean scattering intensity.
The main orientation of the fibers in the bottom part of the sample is clearly recognizable. In the top part of the sample, the fibers are mainly oriented perpendicularly to the plan of the image, with some bundles escaping upwards at the borders. This information is reproduced by a high mean scattering (Fig.~\ref{fig:fig1}c and high brightness in Fig.~\ref{fig:fig1}d), but low fractional anisotropy indicated by the low saturation in Fig.~\ref{fig:fig1}d (color tending to white).

\begin{figure}[h!]
    \centering
    \includegraphics[width=\linewidth]{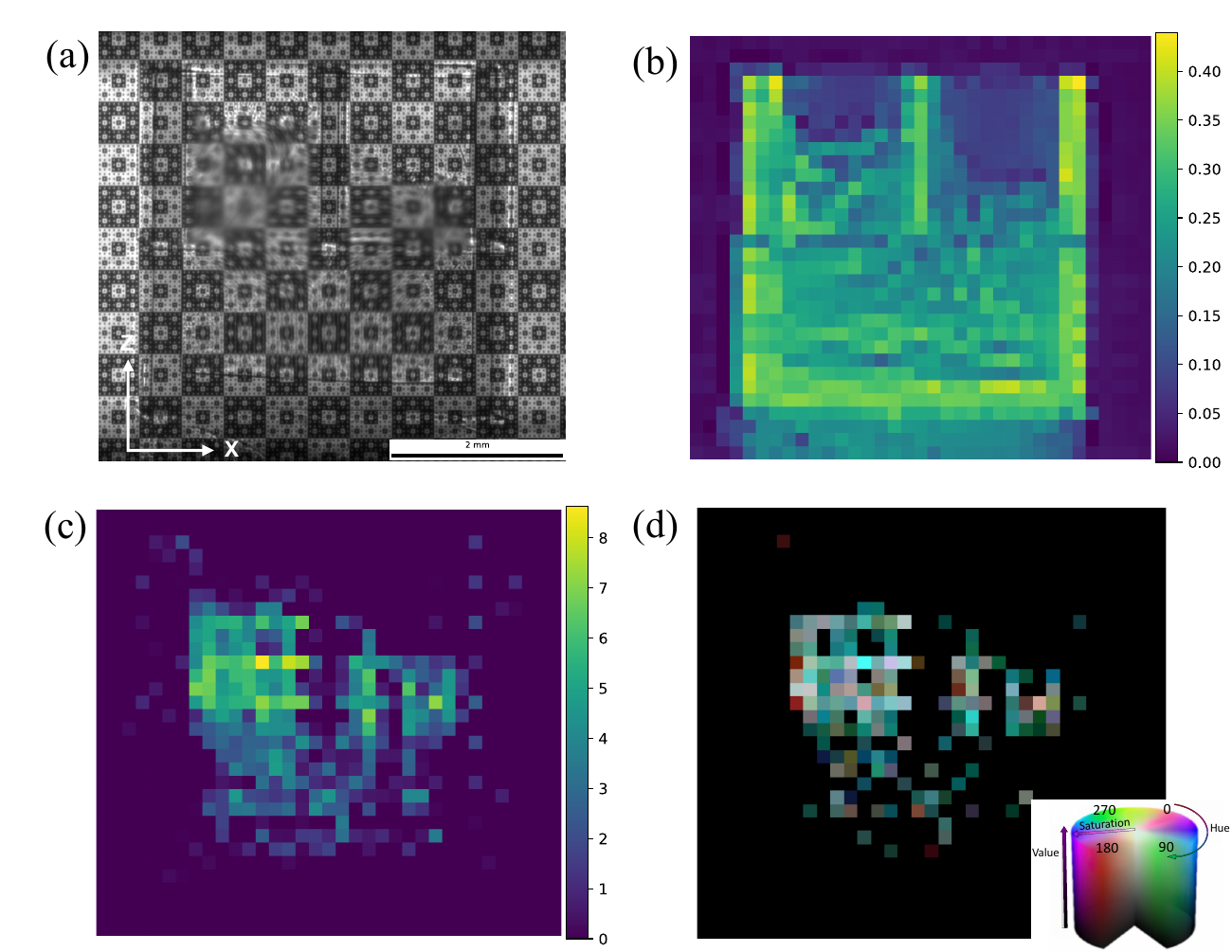}
    \caption{(a) Raw transmission image of the validation sample with fractal array at $\alpha = \SI{0}{\degree}$,  $\beta = \SI{0}{\degree}$. (b) Extracted absorption image, (c) mean scattering in arbitrary units and (d) main orientation signals.}
    \label{fig:fig2}
\end{figure}

\subsection{Fractal pattern array}
The second scan was performed using an in-house fabricated fractal pattern array, featuring structures with different shape and size, which allows tuning of the analysis window (i.e., the image spatial resolution) to focus on a desired feature size. The fractal pattern array can be used as an alternative wavefront modulator. For this scan, images were collected using a PCO.edge 5.5 sCMOS camera, coupled with a 1:1 microscope optic giving an effective pixel size of $\SI{6.5}{\micro\metre}$. The exposure time for each projection was $\SI{100}{\milli\second}$ and the FOV was $802 \times 2560$ pixels.
In this case, the analysis window corresponds to a square whose size is $24$ pixels or $\SI{156}{\micro\metre}$.
The results in Fig.~\ref{fig:fig2} demonstrate that, despite the lower visibility of the pattern compared to the grating scan, the proposed method allows for the extraction of both scattering and main orientation signals also for this kind of reference pattern.

\begin{figure}[h!]
    \centering
    \includegraphics[width=\linewidth]{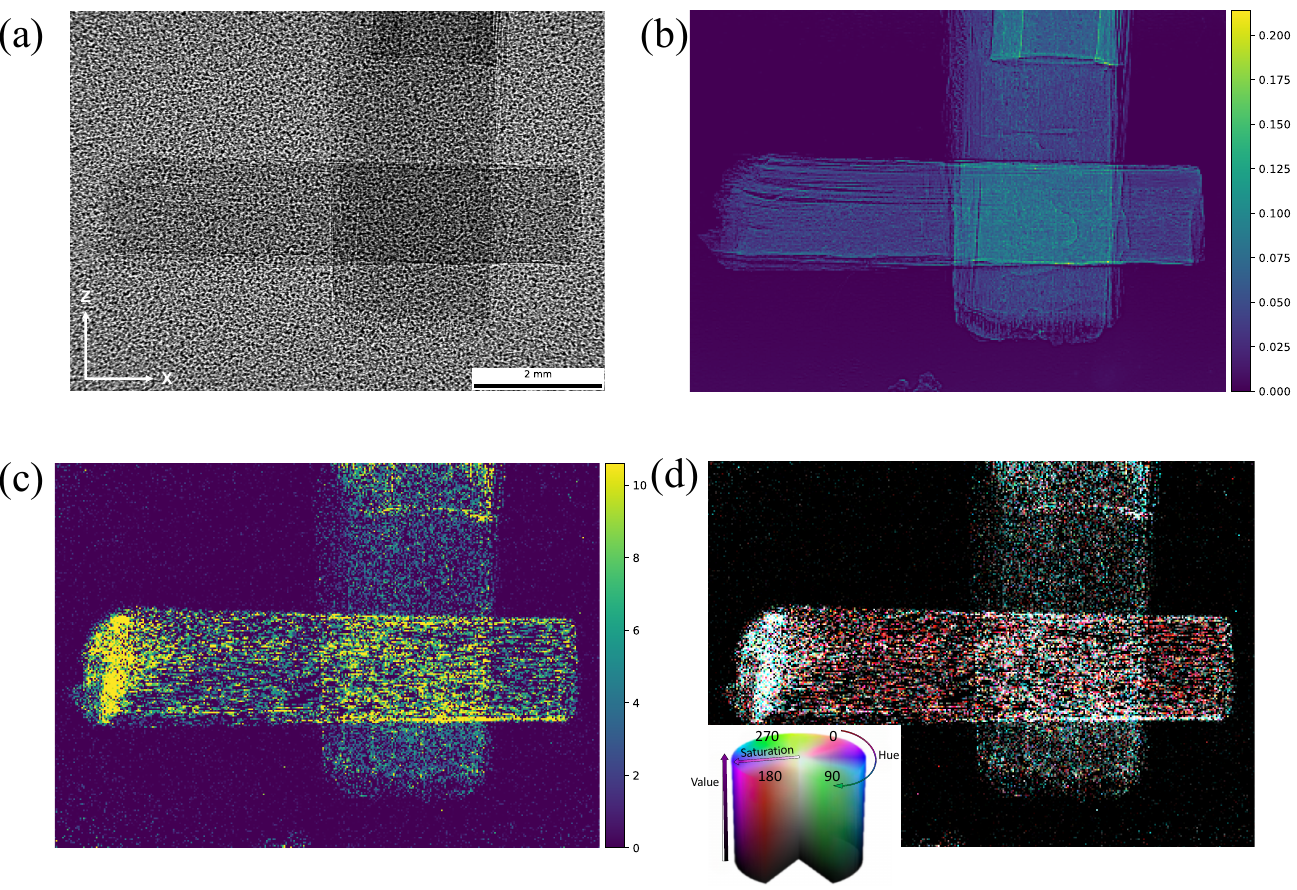}
    \caption{(a) Raw transmission image of a carbon fiber cross with speckles. (b) Extracted absorption image and (c) mean scattering signals. (d) Main orientation signals in green and red, contrast has been enhanced for visualization purposes.}
    \label{fig:fig3}
\end{figure}

\subsection{Random diffuser}
Moving to a setup even easier to implement, sandpaper has proven to be an effective wavefront-marker for dark-field signal extraction.
SBI images using three layers of P320 sandpaper as a wavefront marker were acquired at the SYRMEP beamline (Elettra Sincrotrone Trieste). For this application, the test sample was made of two sheets of unidirectional carbon fibers ($\SI{10}{\micro\metre}$ diameter), glued together to form an angle of $\SI{90}{\degree}$.
We used a filtered white beam with a mean energy of $\SI{37.6}{\kilo\eV}$.
SBI scans were performed using the diffuser-stepping method, where the diffuser is laterally translated ($20$ different positions), while the sample remains stationary, to achieve a higher spatial resolution \cite{DeMarco2023}.
Images were acquired with a water-cooled sCMOS camera (Orca Flash 4.0, Hamamatsu) coupled with an X-ray microscope (Optique Peter). The optics was adjusted to have an effective pixel size of $\SI{3.82}{\micro\metre}$. The detector was placed at a distance of $\SI{52}{\cm}$ from the diffuser, while the diffuser-sample distance was  $\SI{92.5}{\cm}$. The exposure time for each projection was $\SI{75}{\milli\second}$ and the FOV was $1452 \times 2048$ pixels.
Since the pattern created by the sandpaper is random, there is no periodic unit cells as for gratings, hence the analysis window can be of arbitrary size. We chose a square analysis window with size $6$ pixels or $\SI{22.8}{\micro\metre}$.
The results in Fig.~\ref{fig:fig3}d show that the main orientation signal matches with the expected orientation of the fibers (blue along the $y$-axis and purple at $\SI{45}{\degree}$), thus demonstrating the effectiveness of the proposed method for random reference patterns as well.

\subsection{Tensor tomography reconstruction}
\begin{figure}[h!]
    \centering
    \includegraphics[width=\linewidth]{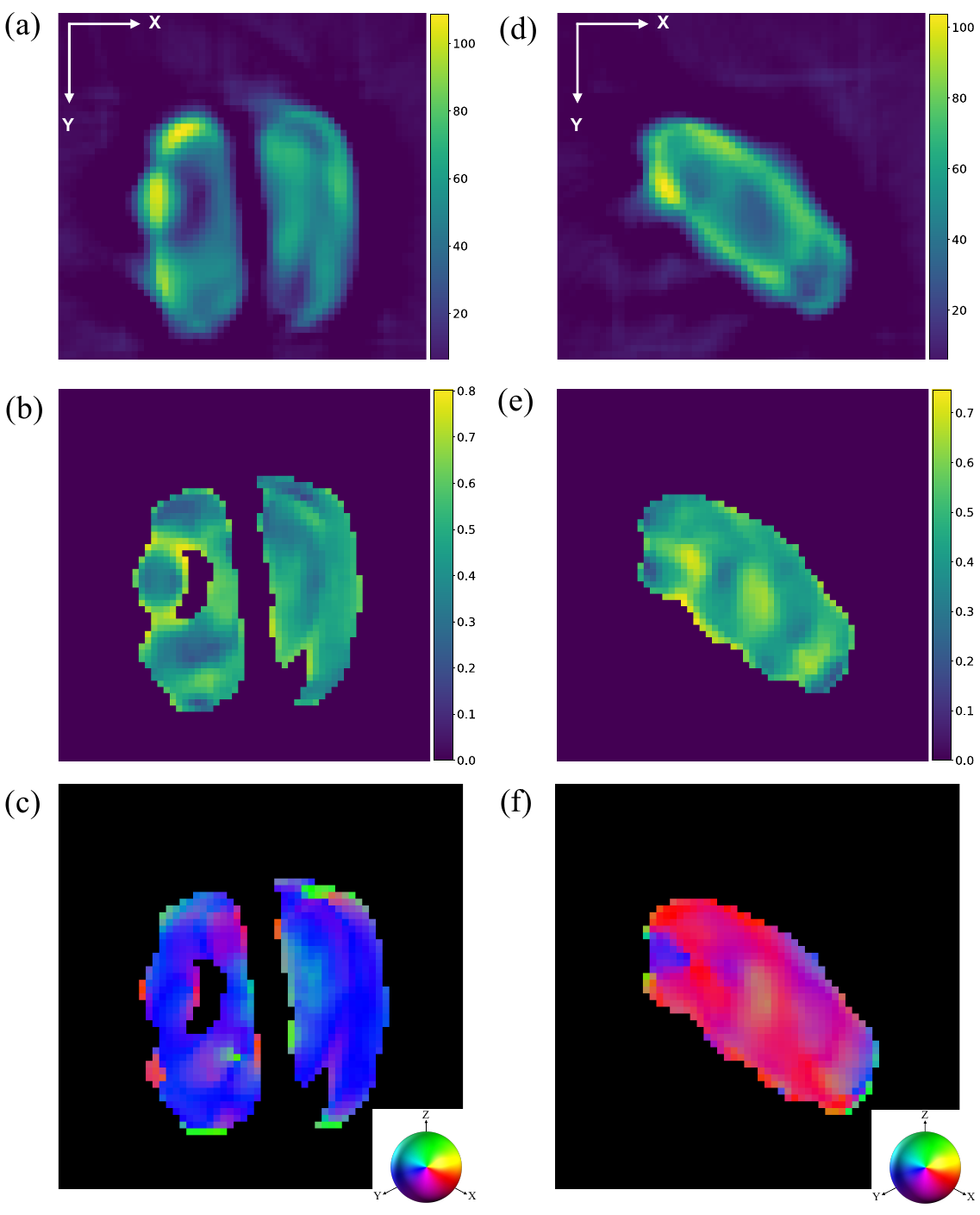}
    \caption{Axial slices through the upper (a-c) and the lower part (d-f) of the validation sample. (a,d) Mean scattering in arbitrary units and (b,e) scattering anisotropy signals in arbitrary units are shown. (c,f) Fiber orientation signals, where the color is an RGB representation of the local structure orientation. The color ball is symmetric with respect to the $x-y$, $x-z$, and $y-z$ planes.}
    \label{fig:fig4}
\end{figure}

Once the omnidirectional dark-field signal is extracted and Eq.~\ref{subeq:A*} is known, it is sufficient to solve the linear system with CG. 
We validated our tensor tomography reconstruction for the circular gratings dataset, in order to compare our results with a previously established method validated only for circular gratings. However, the same analysis can be applied to datasets obtained with the other wavefront markers.
The tomographic slices of the sample are shown in Fig.~\ref{fig:fig4}. Before reconstruction, all projections were aligned using a customized version of the alignment algorithm described in \cite{Liebi2015}. Two axial slices are shown: one through the upper part of the validation sample, and one through the lower part. From the $\mathrm{MS}$ images, it can be seen that in each bundle, there exist clusters with varying local fiber densities rather than a consistent distribution of fibers. The fiber orientation signal images (Figs.~\ref{fig:fig4}c, \ref{fig:fig4}f) show color as an RGB representation of the local structure orientation. The reconstructed main orientations agree  with the orientations of the fibers. However, within each bundle, there are local clusters with orientations that slightly deviate from the main fiber direction. This effect is caused by the fibers placed at the edges, that are bending along the wall.

\begin{figure}[h!]
    \centering
    \includegraphics[width=\linewidth]{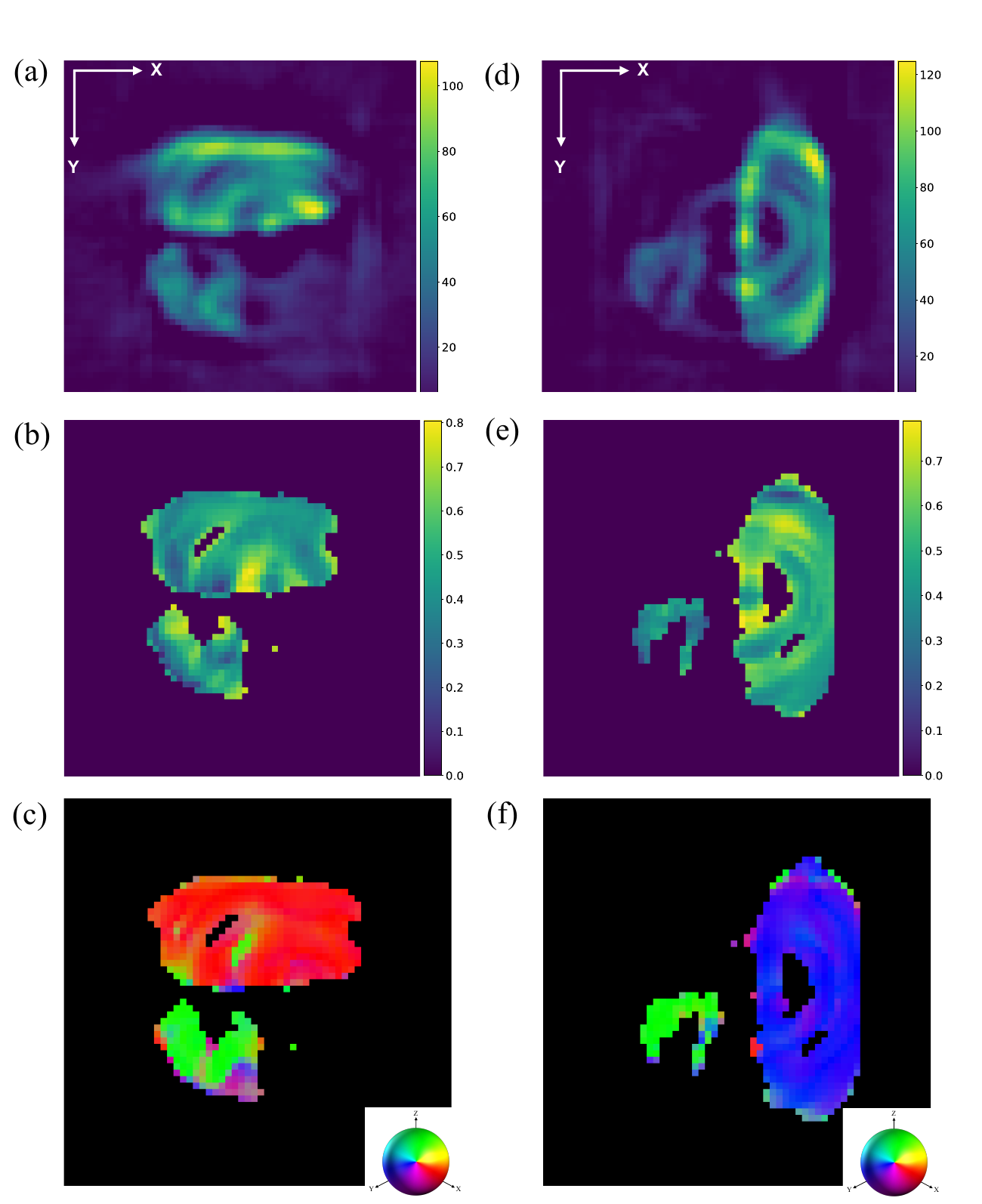}
    \caption{Axial slices through the upper (a-c) and the lower part (d-f) of the validation sample described in \cite{Kim2020}. (a,d) Mean scattering in arbitrary units and (b,e) scattering anisotropy signals in arbitrary units are shown. (c,f) Fiber orientation signals, where the color is an RGB representation of the local structure orientation.}
    \label{fig:fig5}
\end{figure}

We applied our method also to the sample described in \cite{Kim2020}. The only difference with the sample we already described is the orientation of the fibers, which in this case are placed along the three main axes. The fiber orientation signals are shown in Figs.~\ref{fig:fig5}c, \ref{fig:fig5}f. Also in this case the reconstructed main orientations agree with the orientations of the fibers. Local clusters of different orientations are visible also for this sample, as a result of fibers bending along the wall. The reconstructed orientations are comparable with those obtained with the previous method restricted to gratings \cite{Kim2020}.

\begin{figure}[h!]
    \centering
    \includegraphics[width=\linewidth]{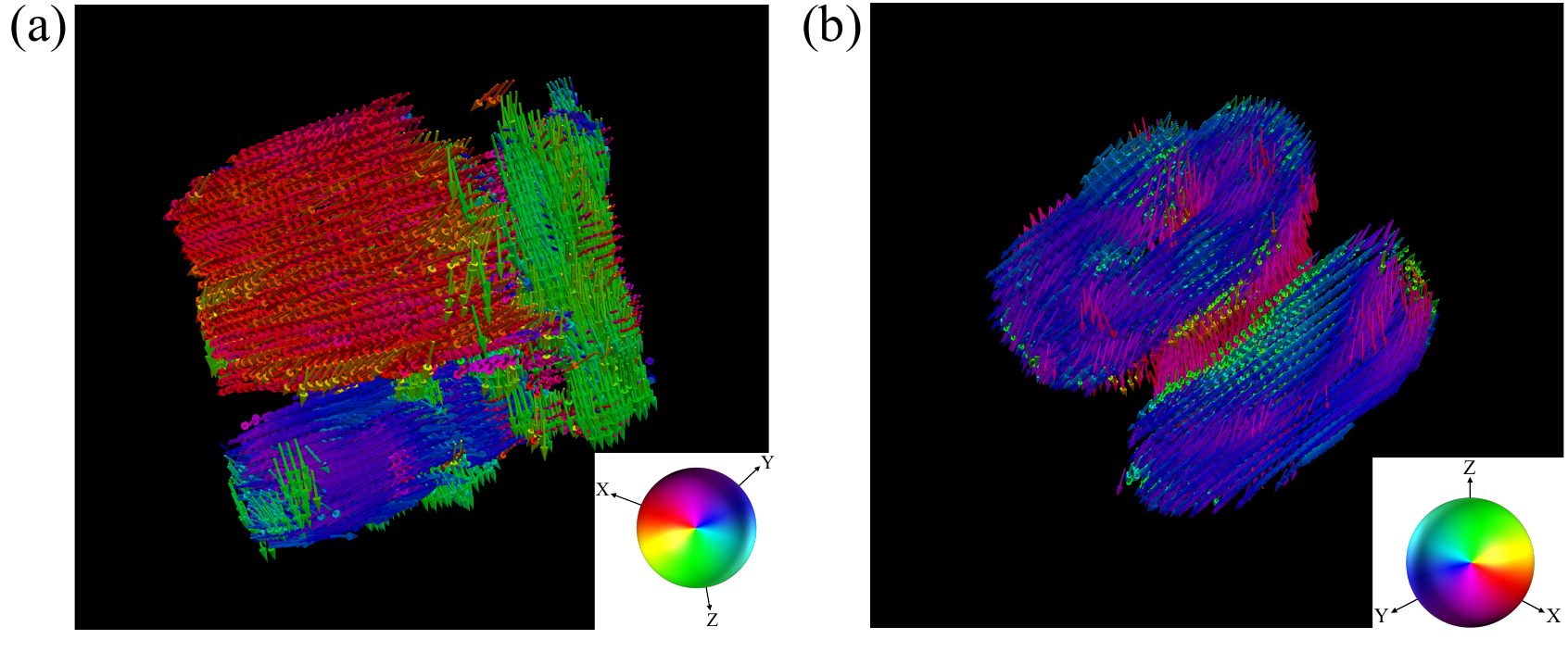}
    \caption{3D visualization of the reconstructed scattering tensor of the two validation samples. In this representation, each arrow's orientation corresponds to the main direction in each voxel.}
    \label{fig:fig6}
\end{figure}

A 3D visualization of the scattering tensor reconstruction of the validation samples is shown in Fig.~\ref{fig:fig6}. In this visualization, each arrow's direction and color represent the main orientation within its respective voxel. We have therefore validated that our method accurately reproduces structure orientations also in tomographic volumes and that the results, in the case of the circular grating array, are comparable with other reconstruction algorithms.

\section{CONCLUSION}
In conclusion, this study introduces a general algorithm applicable to a wide range of X-ray imaging approaches that have access to the scattering signal, without the need for additional regularization. 
We have demonstrated its effectiveness for three different full-field dark-field imaging setups (gratings, fractal, and speckle). Then, we validated the results obtained with GI, comparing them with previous findings. The algorithm has enabled us to pave the way for tensor tomography using SBI, as well as using a fractal wavefront modulator. Our method allows for full reconstruction of tensor tomograms, with a simpler setup and simpler optics. Therefore, it has the potential to transition to laboratory setups, extending its reach to a wider user community. 
The main difference with previous works is the fact that we present a full-field method that can be applied to different wavefront modulators, also non-periodic. The scattering signal extraction and the tensor tomographic reconstruction are independent of the experimental geometry used in previous works. Moreover, we reconstruct the full tensor field, not only the scattering signal along predefined directions.
Several disciplines can profit from this work, to mention only a few: medical physics to study the micro-structural architectures of bones, and material physics to characterize fiber-reinforced materials. The code is available upon reasonable request. 

\section{ACKNOWLEDGEMENTS}
\begin{acknowledgments}
We wish to acknowledge Swiss Light Source (Paul Scherrer Institut) for providing access to its synchrotron radiation facilities and we thank Dr. Christian Schlepuetz and Philipp Zuppiger for assistance during the beamtime. 
Parts of this research were carried out at Elettra Sincrotrone Trieste and we would like to thank Dr. Giuliana Tromba and Dr. Adriano Contillo for assistance in using the SYRMEP beamline.
This publication is part of a project that has received funding from the European Research Council (ERC) under the European Union’s Horizon 2020 research and innovation program (Grant agreement No. 866026).
\end{acknowledgments}

\nocite{*}
\clearpage

\begin{thebibliography}{38}%
	\makeatletter
	\providecommand \@ifxundefined [1]{%
		\@ifx{#1\undefined}
	}%
	\providecommand \@ifnum [1]{%
		\ifnum #1\expandafter \@firstoftwo
		\else \expandafter \@secondoftwo
		\fi
	}%
	\providecommand \@ifx [1]{%
		\ifx #1\expandafter \@firstoftwo
		\else \expandafter \@secondoftwo
		\fi
	}%
	\providecommand \natexlab [1]{#1}%
	\providecommand \enquote  [1]{``#1''}%
	\providecommand \bibnamefont  [1]{#1}%
	\providecommand \bibfnamefont [1]{#1}%
	\providecommand \citenamefont [1]{#1}%
	\providecommand \href@noop [0]{\@secondoftwo}%
	\providecommand \href [0]{\begingroup \@sanitize@url \@href}%
	\providecommand \@href[1]{\@@startlink{#1}\@@href}%
	\providecommand \@@href[1]{\endgroup#1\@@endlink}%
	\providecommand \@sanitize@url [0]{\catcode `\\12\catcode `\$12\catcode
		`\&12\catcode `\#12\catcode `\^12\catcode `\_12\catcode `\%12\relax}%
	\providecommand \@@startlink[1]{}%
	\providecommand \@@endlink[0]{}%
	\providecommand \url  [0]{\begingroup\@sanitize@url \@url }%
	\providecommand \@url [1]{\endgroup\@href {#1}{\urlprefix }}%
	\providecommand \urlprefix  [0]{URL }%
	\providecommand \Eprint [0]{\href }%
	\providecommand \doibase [0]{https://doi.org/}%
	\providecommand \selectlanguage [0]{\@gobble}%
	\providecommand \bibinfo  [0]{\@secondoftwo}%
	\providecommand \bibfield  [0]{\@secondoftwo}%
	\providecommand \translation [1]{[#1]}%
	\providecommand \BibitemOpen [0]{}%
	\providecommand \bibitemStop [0]{}%
	\providecommand \bibitemNoStop [0]{.\EOS\space}%
	\providecommand \EOS [0]{\spacefactor3000\relax}%
	\providecommand \BibitemShut  [1]{\csname bibitem#1\endcsname}%
	\let\auto@bib@innerbib\@empty
	\bibitem [{\citenamefont {Fratzl}(2008)}]{Fratzl2008}%
	\BibitemOpen
	\bibfield  {author} {\bibinfo {author} {\bibfnamefont {P.}~\bibnamefont
			{Fratzl}},\ }\href@noop {} {\emph {\bibinfo {title} {{Collagen: Structure and
					Mechanics, an Introduction}}}}\ (\bibinfo  {publisher} {Springer US},\
	\bibinfo {year} {2008})\BibitemShut {NoStop}%
	\bibitem [{\citenamefont {Byron~P.}\ \emph {et~al.}(1982)\citenamefont
		{Byron~P.}, \citenamefont {McCullough},\ and\ \citenamefont
		{Taggart}}]{Pipes1982}%
	\BibitemOpen
	\bibfield  {author} {\bibinfo {author} {\bibfnamefont {R.}~\bibnamefont
			{Byron~P.}}, \bibinfo {author} {\bibfnamefont {R.~L.}\ \bibnamefont
			{McCullough}},\ and\ \bibinfo {author} {\bibfnamefont {D.~G.}\ \bibnamefont
			{Taggart}},\ }\bibfield  {title} {\bibinfo {title} {{Behavior of
				discontinuous fiber composites: Fiber orientation}},\ }\href@noop {}
	{\bibfield  {journal} {\bibinfo  {journal} {Polymer Composites}\ }\textbf
		{\bibinfo {volume} {3}},\ \bibinfo {pages} {34} (\bibinfo {year}
		{1982})}\BibitemShut {NoStop}%
	\bibitem [{\citenamefont {Schaff}\ \emph {et~al.}(2015)\citenamefont {Schaff},
		\citenamefont {Bech}, \citenamefont {Zaslansky}, \citenamefont {Jud},
		\citenamefont {Liebi}, \citenamefont {Guizar-Sicairos},\ and\ \citenamefont
		{Pfeiffer}}]{Schaff2015}%
	\BibitemOpen
	\bibfield  {author} {\bibinfo {author} {\bibfnamefont {F.}~\bibnamefont
			{Schaff}}, \bibinfo {author} {\bibfnamefont {M.}~\bibnamefont {Bech}},
		\bibinfo {author} {\bibfnamefont {P.}~\bibnamefont {Zaslansky}}, \bibinfo
		{author} {\bibfnamefont {C.}~\bibnamefont {Jud}}, \bibinfo {author}
		{\bibfnamefont {M.}~\bibnamefont {Liebi}}, \bibinfo {author} {\bibfnamefont
			{M.}~\bibnamefont {Guizar-Sicairos}},\ and\ \bibinfo {author} {\bibfnamefont
			{F.}~\bibnamefont {Pfeiffer}},\ }\bibfield  {title} {\bibinfo {title}
		{Six-dimensional real and reciprocal space small-angle {X-ray} scattering
			tomography},\ }\href@noop {} {\bibfield  {journal} {\bibinfo  {journal}
			{Nature}\ }\textbf {\bibinfo {volume} {527}},\ \bibinfo {pages} {353}
		(\bibinfo {year} {2015})}\BibitemShut {NoStop}%
	\bibitem [{\citenamefont {Liebi}\ \emph {et~al.}(2015)\citenamefont {Liebi},
		\citenamefont {Georgiadis}, \citenamefont {Menzel}, \citenamefont
		{Schneider}, \citenamefont {Kohlbrecher}, \citenamefont {Bunk},\ and\
		\citenamefont {Guizar-Sicairos}}]{Liebi2015}%
	\BibitemOpen
	\bibfield  {author} {\bibinfo {author} {\bibfnamefont {M.}~\bibnamefont
			{Liebi}}, \bibinfo {author} {\bibfnamefont {M.}~\bibnamefont {Georgiadis}},
		\bibinfo {author} {\bibfnamefont {A.}~\bibnamefont {Menzel}}, \bibinfo
		{author} {\bibfnamefont {P.}~\bibnamefont {Schneider}}, \bibinfo {author}
		{\bibfnamefont {J.}~\bibnamefont {Kohlbrecher}}, \bibinfo {author}
		{\bibfnamefont {O.}~\bibnamefont {Bunk}},\ and\ \bibinfo {author}
		{\bibfnamefont {M.}~\bibnamefont {Guizar-Sicairos}},\ }\bibfield  {title}
	{\bibinfo {title} {Nanostructure surveys of macroscopic specimens by
			small-angle scattering tensor tomography},\ }\href@noop {} {\bibfield
		{journal} {\bibinfo  {journal} {Nature}\ }\textbf {\bibinfo {volume} {527}},\
		\bibinfo {pages} {349} (\bibinfo {year} {2015})}\BibitemShut {NoStop}%
	\bibitem [{\citenamefont {Liebi}\ \emph {et~al.}(2018)\citenamefont {Liebi},
		\citenamefont {Georgiadis}, \citenamefont {Kohlbrecher}, \citenamefont
		{Holler}, \citenamefont {Raabe}, \citenamefont {Usov}, \citenamefont
		{Menzel}, \citenamefont {Schneider}, \citenamefont {Bunk},\ and\
		\citenamefont {Guizar-Sicairos}}]{Liebi2018}%
	\BibitemOpen
	\bibfield  {author} {\bibinfo {author} {\bibfnamefont {M.}~\bibnamefont
			{Liebi}}, \bibinfo {author} {\bibfnamefont {M.}~\bibnamefont {Georgiadis}},
		\bibinfo {author} {\bibfnamefont {J.}~\bibnamefont {Kohlbrecher}}, \bibinfo
		{author} {\bibfnamefont {M.}~\bibnamefont {Holler}}, \bibinfo {author}
		{\bibfnamefont {J.}~\bibnamefont {Raabe}}, \bibinfo {author} {\bibfnamefont
			{I.}~\bibnamefont {Usov}}, \bibinfo {author} {\bibfnamefont {A.}~\bibnamefont
			{Menzel}}, \bibinfo {author} {\bibfnamefont {P.}~\bibnamefont {Schneider}},
		\bibinfo {author} {\bibfnamefont {O.}~\bibnamefont {Bunk}},\ and\ \bibinfo
		{author} {\bibfnamefont {M.}~\bibnamefont {Guizar-Sicairos}},\ }\bibfield
	{title} {\bibinfo {title} {{Small-angle X-ray scattering tensor tomography:
				model of the three-dimensional reciprocal-space map, reconstruction algorithm
				and angular sampling requirements}},\ }\href@noop {} {\bibfield  {journal}
		{\bibinfo  {journal} {Acta Crystallographica Section A}\ }\textbf {\bibinfo
			{volume} {74}},\ \bibinfo {pages} {12} (\bibinfo {year} {2018})}\BibitemShut
	{NoStop}%
	\bibitem [{\citenamefont {Gao}\ \emph {et~al.}(2019)\citenamefont {Gao},
		\citenamefont {Guizar-Sicairos}, \citenamefont {Lutz-Bueno}, \citenamefont
		{Schr{\"{o}}ter}, \citenamefont {Liebi}, \citenamefont {Rudin},\ and\
		\citenamefont {Georgiadis}}]{Gao2019}%
	\BibitemOpen
	\bibfield  {author} {\bibinfo {author} {\bibfnamefont {Z.}~\bibnamefont
			{Gao}}, \bibinfo {author} {\bibfnamefont {M.}~\bibnamefont
			{Guizar-Sicairos}}, \bibinfo {author} {\bibfnamefont {V.}~\bibnamefont
			{Lutz-Bueno}}, \bibinfo {author} {\bibfnamefont {A.}~\bibnamefont
			{Schr{\"{o}}ter}}, \bibinfo {author} {\bibfnamefont {M.}~\bibnamefont
			{Liebi}}, \bibinfo {author} {\bibfnamefont {M.}~\bibnamefont {Rudin}},\ and\
		\bibinfo {author} {\bibfnamefont {M.}~\bibnamefont {Georgiadis}},\ }\bibfield
	{title} {\bibinfo {title} {High-speed tensor tomography: iterative
			reconstruction tensor tomography {(IRTT)} algorithm},\ }\href@noop {}
	{\bibfield  {journal} {\bibinfo  {journal} {Acta Crystallographica Section
				A}\ }\textbf {\bibinfo {volume} {75}},\ \bibinfo {pages} {223} (\bibinfo
		{year} {2019})}\BibitemShut {NoStop}%
	\bibitem [{\citenamefont {Skj{\o}nsfjell}\ \emph {et~al.}(2016)\citenamefont
		{Skj{\o}nsfjell}, \citenamefont {Kringeland}, \citenamefont {Granlund},
		\citenamefont {H{\o}ydalsvik}, \citenamefont {Diaz},\ and\ \citenamefont
		{Breiby}}]{Skjonsfjell2016}%
	\BibitemOpen
	\bibfield  {author} {\bibinfo {author} {\bibfnamefont {E.~T.}\ \bibnamefont
			{Skj{\o}nsfjell}}, \bibinfo {author} {\bibfnamefont {T.}~\bibnamefont
			{Kringeland}}, \bibinfo {author} {\bibfnamefont {H.}~\bibnamefont
			{Granlund}}, \bibinfo {author} {\bibfnamefont {K.}~\bibnamefont
			{H{\o}ydalsvik}}, \bibinfo {author} {\bibfnamefont {A.}~\bibnamefont
			{Diaz}},\ and\ \bibinfo {author} {\bibfnamefont {D.~W.}\ \bibnamefont
			{Breiby}},\ }\bibfield  {title} {\bibinfo {title} {{Retrieving the spatially
				resolved preferred orientation of embedded anisotropic particles by
				small-angle X-ray scattering tomography}},\ }\href@noop {} {\bibfield
		{journal} {\bibinfo  {journal} {Journal of Applied Crystallography}\ }\textbf
		{\bibinfo {volume} {49}},\ \bibinfo {pages} {902} (\bibinfo {year}
		{2016})}\BibitemShut {NoStop}%
	\bibitem [{\citenamefont {Momose}\ \emph {et~al.}(2003)\citenamefont {Momose},
		\citenamefont {Kawamoto}, \citenamefont {Koyama}, \citenamefont {Hamaishi},
		\citenamefont {Takai},\ and\ \citenamefont {Suzuki}}]{Momose2003}%
	\BibitemOpen
	\bibfield  {author} {\bibinfo {author} {\bibfnamefont {A.}~\bibnamefont
			{Momose}}, \bibinfo {author} {\bibfnamefont {S.}~\bibnamefont {Kawamoto}},
		\bibinfo {author} {\bibfnamefont {I.}~\bibnamefont {Koyama}}, \bibinfo
		{author} {\bibfnamefont {Y.}~\bibnamefont {Hamaishi}}, \bibinfo {author}
		{\bibfnamefont {K.}~\bibnamefont {Takai}},\ and\ \bibinfo {author}
		{\bibfnamefont {Y.}~\bibnamefont {Suzuki}},\ }\bibfield  {title} {\bibinfo
		{title} {{Demonstration of X-Ray Talbot Interferometry}},\ }\href@noop {}
	{\bibfield  {journal} {\bibinfo  {journal} {Japanese Journal of Applied
				Physics}\ }\textbf {\bibinfo {volume} {42}},\ \bibinfo {pages} {L866}
		(\bibinfo {year} {2003})}\BibitemShut {NoStop}%
	\bibitem [{\citenamefont {Weitkamp}\ \emph {et~al.}(2005)\citenamefont
		{Weitkamp}, \citenamefont {Diaz}, \citenamefont {David}, \citenamefont
		{Pfeiffer}, \citenamefont {Stampanoni}, \citenamefont {Cloetens},\ and\
		\citenamefont {Ziegler}}]{Weitkamp2005}%
	\BibitemOpen
	\bibfield  {author} {\bibinfo {author} {\bibfnamefont {T.}~\bibnamefont
			{Weitkamp}}, \bibinfo {author} {\bibfnamefont {A.}~\bibnamefont {Diaz}},
		\bibinfo {author} {\bibfnamefont {C.}~\bibnamefont {David}}, \bibinfo
		{author} {\bibfnamefont {F.}~\bibnamefont {Pfeiffer}}, \bibinfo {author}
		{\bibfnamefont {M.}~\bibnamefont {Stampanoni}}, \bibinfo {author}
		{\bibfnamefont {P.}~\bibnamefont {Cloetens}},\ and\ \bibinfo {author}
		{\bibfnamefont {E.}~\bibnamefont {Ziegler}},\ }\bibfield  {title} {\bibinfo
		{title} {{X-ray} phase imaging with a grating interferometer},\ }\href@noop
	{} {\bibfield  {journal} {\bibinfo  {journal} {Optics Express}\ }\textbf
		{\bibinfo {volume} {13}},\ \bibinfo {pages} {6296} (\bibinfo {year}
		{2005})}\BibitemShut {NoStop}%
	\bibitem [{\citenamefont {Pfeiffer}\ \emph {et~al.}(2008)\citenamefont
		{Pfeiffer}, \citenamefont {Bech}, \citenamefont {Bunk}, \citenamefont
		{Kraft}, \citenamefont {Eikenberry}, \citenamefont {Brönnimann},
		\citenamefont {Grünzweig},\ and\ \citenamefont {David}}]{Pfeiffer2008}%
	\BibitemOpen
	\bibfield  {author} {\bibinfo {author} {\bibfnamefont {F.}~\bibnamefont
			{Pfeiffer}}, \bibinfo {author} {\bibfnamefont {M.}~\bibnamefont {Bech}},
		\bibinfo {author} {\bibfnamefont {O.}~\bibnamefont {Bunk}}, \bibinfo {author}
		{\bibfnamefont {P.}~\bibnamefont {Kraft}}, \bibinfo {author} {\bibfnamefont
			{E.~F.}\ \bibnamefont {Eikenberry}}, \bibinfo {author} {\bibfnamefont
			{C.}~\bibnamefont {Brönnimann}}, \bibinfo {author} {\bibfnamefont
			{C.}~\bibnamefont {Grünzweig}},\ and\ \bibinfo {author} {\bibfnamefont
			{C.}~\bibnamefont {David}},\ }\bibfield  {title} {\bibinfo {title}
		{Hard-x-ray dark-field imaging using a grating interferometer},\ }\href@noop
	{} {\bibfield  {journal} {\bibinfo  {journal} {Nature Materials}\ }\textbf
		{\bibinfo {volume} {7}},\ \bibinfo {pages} {134} (\bibinfo {year}
		{2008})}\BibitemShut {NoStop}%
	\bibitem [{\citenamefont {Morgan}\ \emph {et~al.}(2012)\citenamefont {Morgan},
		\citenamefont {Paganin},\ and\ \citenamefont {Siu}}]{Morgan2012}%
	\BibitemOpen
	\bibfield  {author} {\bibinfo {author} {\bibfnamefont {K.~S.}\ \bibnamefont
			{Morgan}}, \bibinfo {author} {\bibfnamefont {D.~M.}\ \bibnamefont
			{Paganin}},\ and\ \bibinfo {author} {\bibfnamefont {K.~K.~W.}\ \bibnamefont
			{Siu}},\ }\bibfield  {title} {\bibinfo {title} {{X-ray} phase imaging with a
			paper analyzer},\ }\href@noop {} {\bibfield  {journal} {\bibinfo  {journal}
			{Applied Physics Letters}\ }\textbf {\bibinfo {volume} {100}},\ \bibinfo
		{pages} {124102} (\bibinfo {year} {2012})}\BibitemShut {NoStop}%
	\bibitem [{\citenamefont {Berujon}\ \emph {et~al.}(2012)\citenamefont
		{Berujon}, \citenamefont {Wang},\ and\ \citenamefont
		{Sawhney}}]{Berujon2012}%
	\BibitemOpen
	\bibfield  {author} {\bibinfo {author} {\bibfnamefont {S.}~\bibnamefont
			{Berujon}}, \bibinfo {author} {\bibfnamefont {H.}~\bibnamefont {Wang}},\ and\
		\bibinfo {author} {\bibfnamefont {K.}~\bibnamefont {Sawhney}},\ }\bibfield
	{title} {\bibinfo {title} {{X-ray} multimodal imaging using a random-phase
			object},\ }\href@noop {} {\bibfield  {journal} {\bibinfo  {journal} {Physical
				Review A}\ }\textbf {\bibinfo {volume} {86}},\ \bibinfo {pages} {063813}
		(\bibinfo {year} {2012})}\BibitemShut {NoStop}%
	\bibitem [{\citenamefont {Zanette}\ \emph {et~al.}(2015)\citenamefont
		{Zanette}, \citenamefont {Zdora}, \citenamefont {T.~Zhou}, \citenamefont
		{Larsson}, \citenamefont {Thibault}, \citenamefont {Hertz},\ and\
		\citenamefont {Pfeiffer}}]{Zanette2015}%
	\BibitemOpen
	\bibfield  {author} {\bibinfo {author} {\bibfnamefont {I.}~\bibnamefont
			{Zanette}}, \bibinfo {author} {\bibfnamefont {M.~C.}\ \bibnamefont {Zdora}},
		\bibinfo {author} {\bibfnamefont {A.~B.}\ \bibnamefont {T.~Zhou}}, \bibinfo
		{author} {\bibfnamefont {D.~H.}\ \bibnamefont {Larsson}}, \bibinfo {author}
		{\bibfnamefont {P.}~\bibnamefont {Thibault}}, \bibinfo {author}
		{\bibfnamefont {H.}~\bibnamefont {Hertz}},\ and\ \bibinfo {author}
		{\bibfnamefont {F.}~\bibnamefont {Pfeiffer}},\ }\bibfield  {title} {\bibinfo
		{title} {{X-ray} microtomography using correlation of near-field speckles for
			material characterization},\ }\href@noop {} {\bibfield  {journal} {\bibinfo
			{journal} {Proceedings of the National Academy of Sciences of the United
				States of America}\ }\textbf {\bibinfo {volume} {112}},\ \bibinfo {pages}
		{12569} (\bibinfo {year} {2015})}\BibitemShut {NoStop}%
	\bibitem [{\citenamefont {Zdora}(2018)}]{Zdora2018}%
	\BibitemOpen
	\bibfield  {author} {\bibinfo {author} {\bibfnamefont {M.-C.}\ \bibnamefont
			{Zdora}},\ }\bibfield  {title} {\bibinfo {title} {{State of the Art of X-ray
				Speckle-Based Phase-Contrast and Dark-Field Imaging}},\ }\href@noop {}
	{\bibfield  {journal} {\bibinfo  {journal} {Journal of Imaging}\ }\textbf
		{\bibinfo {volume} {4}},\ \bibinfo {pages} {60} (\bibinfo {year}
		{2018})}\BibitemShut {NoStop}%
	\bibitem [{\citenamefont {Zhou}\ \emph {et~al.}(2018)\citenamefont {Zhou},
		\citenamefont {Wang},\ and\ \citenamefont {Sawhney}}]{Zhou2018}%
	\BibitemOpen
	\bibfield  {author} {\bibinfo {author} {\bibfnamefont {T.}~\bibnamefont
			{Zhou}}, \bibinfo {author} {\bibfnamefont {H.}~\bibnamefont {Wang}},\ and\
		\bibinfo {author} {\bibfnamefont {K.}~\bibnamefont {Sawhney}},\ }\bibfield
	{title} {\bibinfo {title} {Single-shot {X-ray} dark-field imaging with
			omnidirectional sensitivity using random-pattern wavefront modulator},\
	}\href@noop {} {\bibfield  {journal} {\bibinfo  {journal} {Applied Physics
				Letters}\ }\textbf {\bibinfo {volume} {113}},\ \bibinfo {pages} {091102}
		(\bibinfo {year} {2018})}\BibitemShut {NoStop}%
	\bibitem [{\citenamefont {Zdora}(2021)}]{Zdora2021}%
	\BibitemOpen
	\bibfield  {author} {\bibinfo {author} {\bibfnamefont {M.~C.}\ \bibnamefont
			{Zdora}},\ }\href@noop {} {\emph {\bibinfo {title} {{X-ray} phase-contrast
				imaging using near-field speckles}}}\ (\bibinfo  {publisher} {Springer
		Cham},\ \bibinfo {year} {2021})\BibitemShut {NoStop}%
	\bibitem [{\citenamefont {Yashiro}\ \emph {et~al.}(2010)\citenamefont
		{Yashiro}, \citenamefont {Terui}, \citenamefont {Kawabata},\ and\
		\citenamefont {Momose}}]{Yashiro2010}%
	\BibitemOpen
	\bibfield  {author} {\bibinfo {author} {\bibfnamefont {W.}~\bibnamefont
			{Yashiro}}, \bibinfo {author} {\bibfnamefont {Y.}~\bibnamefont {Terui}},
		\bibinfo {author} {\bibfnamefont {K.}~\bibnamefont {Kawabata}},\ and\
		\bibinfo {author} {\bibfnamefont {A.}~\bibnamefont {Momose}},\ }\bibfield
	{title} {\bibinfo {title} {{On the origin of visibility contrast in x-ray
				Talbot interferometry}},\ }\href@noop {} {\bibfield  {journal} {\bibinfo
			{journal} {Optics Express}\ }\textbf {\bibinfo {volume} {18}},\ \bibinfo
		{pages} {16890} (\bibinfo {year} {2010})}\BibitemShut {NoStop}%
	\bibitem [{\citenamefont {Lynch}\ \emph {et~al.}(2011)\citenamefont {Lynch},
		\citenamefont {Pai}, \citenamefont {Auxier}, \citenamefont {Stein},
		\citenamefont {Bennett}, \citenamefont {Kemble}, \citenamefont {Xiao},
		\citenamefont {Lee}, \citenamefont {Morgan},\ and\ \citenamefont
		{Wen}}]{Lynch2011}%
	\BibitemOpen
	\bibfield  {author} {\bibinfo {author} {\bibfnamefont {S.~K.}\ \bibnamefont
			{Lynch}}, \bibinfo {author} {\bibfnamefont {V.}~\bibnamefont {Pai}}, \bibinfo
		{author} {\bibfnamefont {J.}~\bibnamefont {Auxier}}, \bibinfo {author}
		{\bibfnamefont {A.~F.}\ \bibnamefont {Stein}}, \bibinfo {author}
		{\bibfnamefont {E.~E.}\ \bibnamefont {Bennett}}, \bibinfo {author}
		{\bibfnamefont {C.~K.}\ \bibnamefont {Kemble}}, \bibinfo {author}
		{\bibfnamefont {X.}~\bibnamefont {Xiao}}, \bibinfo {author} {\bibfnamefont
			{W.-K.}\ \bibnamefont {Lee}}, \bibinfo {author} {\bibfnamefont {N.~Y.}\
			\bibnamefont {Morgan}},\ and\ \bibinfo {author} {\bibfnamefont {H.~H.}\
			\bibnamefont {Wen}},\ }\bibfield  {title} {\bibinfo {title} {Interpretation
			of dark-field contrast and particle-size selectivity in grating
			interferometers},\ }\href@noop {} {\bibfield  {journal} {\bibinfo  {journal}
			{Applied Optics}\ }\textbf {\bibinfo {volume} {50}},\ \bibinfo {pages} {4310}
		(\bibinfo {year} {2011})}\BibitemShut {NoStop}%
	\bibitem [{\citenamefont {Strobl}(2014)}]{Strobl2014}%
	\BibitemOpen
	\bibfield  {author} {\bibinfo {author} {\bibfnamefont {M.}~\bibnamefont
			{Strobl}},\ }\bibfield  {title} {\bibinfo {title} {General solution for
			quantitative dark-field contrast imaging with grating interferometers},\
	}\href@noop {} {\bibfield  {journal} {\bibinfo  {journal} {Scientific
				Reports}\ }\textbf {\bibinfo {volume} {4}},\ \bibinfo {pages} {7243}
		(\bibinfo {year} {2014})}\BibitemShut {NoStop}%
	\bibitem [{\citenamefont {Malecki}\ \emph {et~al.}(2014)\citenamefont
		{Malecki}, \citenamefont {Potdevin}, \citenamefont {T.~Biernath},
		\citenamefont {K.~Willer}, \citenamefont {Maisenbacher}, \citenamefont
		{Gibmeier}, \citenamefont {Wanner},\ and\ \citenamefont
		{Pfeiffer}}]{Malecki2014}%
	\BibitemOpen
	\bibfield  {author} {\bibinfo {author} {\bibfnamefont {A.}~\bibnamefont
			{Malecki}}, \bibinfo {author} {\bibfnamefont {G.}~\bibnamefont {Potdevin}},
		\bibinfo {author} {\bibfnamefont {E.~E.}\ \bibnamefont {T.~Biernath}},
		\bibinfo {author} {\bibfnamefont {T.~L.}\ \bibnamefont {K.~Willer}}, \bibinfo
		{author} {\bibfnamefont {J.}~\bibnamefont {Maisenbacher}}, \bibinfo {author}
		{\bibfnamefont {J.}~\bibnamefont {Gibmeier}}, \bibinfo {author}
		{\bibfnamefont {A.}~\bibnamefont {Wanner}},\ and\ \bibinfo {author}
		{\bibfnamefont {F.}~\bibnamefont {Pfeiffer}},\ }\bibfield  {title} {\bibinfo
		{title} {{X-ray} tensor tomography},\ }\href@noop {} {\bibfield  {journal}
		{\bibinfo  {journal} {Europhysics Letters}\ }\textbf {\bibinfo {volume}
			{105}},\ \bibinfo {pages} {38002} (\bibinfo {year} {2014})}\BibitemShut
	{NoStop}%
	\bibitem [{\citenamefont {Vogel}\ \emph {et~al.}(2015)\citenamefont {Vogel},
		\citenamefont {Schaff}, \citenamefont {Fehringer}, \citenamefont {C.~Jud},
		\citenamefont {Pfeiffer},\ and\ \citenamefont {Lasser}}]{Vogel2015}%
	\BibitemOpen
	\bibfield  {author} {\bibinfo {author} {\bibfnamefont {J.}~\bibnamefont
			{Vogel}}, \bibinfo {author} {\bibfnamefont {F.}~\bibnamefont {Schaff}},
		\bibinfo {author} {\bibfnamefont {A.}~\bibnamefont {Fehringer}}, \bibinfo
		{author} {\bibfnamefont {M.~W.}\ \bibnamefont {C.~Jud}}, \bibinfo {author}
		{\bibfnamefont {F.}~\bibnamefont {Pfeiffer}},\ and\ \bibinfo {author}
		{\bibfnamefont {T.}~\bibnamefont {Lasser}},\ }\bibfield  {title} {\bibinfo
		{title} {Constrained {X-ray} tensor tomography reconstruction},\ }\href@noop
	{} {\bibfield  {journal} {\bibinfo  {journal} {Opticts Express}\ }\textbf
		{\bibinfo {volume} {23}},\ \bibinfo {pages} {15134} (\bibinfo {year}
		{2015})}\BibitemShut {NoStop}%
	\bibitem [{\citenamefont {Jensen}\ \emph {et~al.}(2010)\citenamefont {Jensen},
		\citenamefont {Bech}, \citenamefont {Bunk}, \citenamefont {Donath},
		\citenamefont {David}, \citenamefont {Feidenhans'l},\ and\ \citenamefont
		{Pfeiffer}}]{Jensen2010}%
	\BibitemOpen
	\bibfield  {author} {\bibinfo {author} {\bibfnamefont {T.~H.}\ \bibnamefont
			{Jensen}}, \bibinfo {author} {\bibfnamefont {M.}~\bibnamefont {Bech}},
		\bibinfo {author} {\bibfnamefont {O.}~\bibnamefont {Bunk}}, \bibinfo {author}
		{\bibfnamefont {T.}~\bibnamefont {Donath}}, \bibinfo {author} {\bibfnamefont
			{C.}~\bibnamefont {David}}, \bibinfo {author} {\bibfnamefont
			{R.}~\bibnamefont {Feidenhans'l}},\ and\ \bibinfo {author} {\bibfnamefont
			{F.}~\bibnamefont {Pfeiffer}},\ }\bibfield  {title} {\bibinfo {title}
		{Directional x-ray dark-field imaging},\ }\href@noop {} {\bibfield  {journal}
		{\bibinfo  {journal} {Physics in Medicine \& Biology}\ }\textbf {\bibinfo
			{volume} {55}},\ \bibinfo {pages} {3317} (\bibinfo {year}
		{2010})}\BibitemShut {NoStop}%
	\bibitem [{\citenamefont {Kim}\ \emph {et~al.}(2022{\natexlab{a}})\citenamefont
		{Kim}, \citenamefont {Slyamov}, \citenamefont {Lauridsen}, \citenamefont
		{Birkbak}, \citenamefont {Ramos}, \citenamefont {Marone}, \citenamefont
		{Andreasen}, \citenamefont {Stampanoni},\ and\ \citenamefont
		{Kagias}}]{Kim2022b}%
	\BibitemOpen
	\bibfield  {author} {\bibinfo {author} {\bibfnamefont {J.}~\bibnamefont
			{Kim}}, \bibinfo {author} {\bibfnamefont {A.}~\bibnamefont {Slyamov}},
		\bibinfo {author} {\bibfnamefont {E.}~\bibnamefont {Lauridsen}}, \bibinfo
		{author} {\bibfnamefont {M.}~\bibnamefont {Birkbak}}, \bibinfo {author}
		{\bibfnamefont {T.}~\bibnamefont {Ramos}}, \bibinfo {author} {\bibfnamefont
			{F.}~\bibnamefont {Marone}}, \bibinfo {author} {\bibfnamefont {J.~W.}\
			\bibnamefont {Andreasen}}, \bibinfo {author} {\bibfnamefont {M.}~\bibnamefont
			{Stampanoni}},\ and\ \bibinfo {author} {\bibfnamefont {M.}~\bibnamefont
			{Kagias}},\ }\bibfield  {title} {\bibinfo {title} {Macroscopic mapping of
			microscale fibers in freeform injection molded fiber-reinforced composites
			using {X-ray} scattering tensor tomography},\ }\href@noop {} {\bibfield
		{journal} {\bibinfo  {journal} {Composites Part B: Engineering}\ }\textbf
		{\bibinfo {volume} {233}},\ \bibinfo {pages} {109634} (\bibinfo {year}
		{2022}{\natexlab{a}})}\BibitemShut {NoStop}%
	\bibitem [{\citenamefont {Smith}\ \emph {et~al.}(2022)\citenamefont {Smith},
		\citenamefont {De~Marco}, \citenamefont {Broche}, \citenamefont {Zdora},
		\citenamefont {Phillips}, \citenamefont {Boardman},\ and\ \citenamefont
		{Thibault}}]{Smith2022}%
	\BibitemOpen
	\bibfield  {author} {\bibinfo {author} {\bibfnamefont {R.}~\bibnamefont
			{Smith}}, \bibinfo {author} {\bibfnamefont {F.}~\bibnamefont {De~Marco}},
		\bibinfo {author} {\bibfnamefont {L.}~\bibnamefont {Broche}}, \bibinfo
		{author} {\bibfnamefont {M.-C.}\ \bibnamefont {Zdora}}, \bibinfo {author}
		{\bibfnamefont {N.~W.}\ \bibnamefont {Phillips}}, \bibinfo {author}
		{\bibfnamefont {R.}~\bibnamefont {Boardman}},\ and\ \bibinfo {author}
		{\bibfnamefont {P.}~\bibnamefont {Thibault}},\ }\bibfield  {title} {\bibinfo
		{title} {{X-ray directional dark-field imaging using Unified Modulated
				Pattern Analysis}},\ }\href@noop {} {\bibfield  {journal} {\bibinfo
			{journal} {PLOS ONE}\ }\textbf {\bibinfo {volume} {17}},\ \bibinfo {pages}
		{e0273315} (\bibinfo {year} {2022})}\BibitemShut {NoStop}%
	\bibitem [{\citenamefont {Kim}\ \emph {et~al.}(2020)\citenamefont {Kim},
		\citenamefont {Kagias}, \citenamefont {Marone},\ and\ \citenamefont
		{Stampanoni}}]{Kim2020}%
	\BibitemOpen
	\bibfield  {author} {\bibinfo {author} {\bibfnamefont {J.}~\bibnamefont
			{Kim}}, \bibinfo {author} {\bibfnamefont {M.}~\bibnamefont {Kagias}},
		\bibinfo {author} {\bibfnamefont {F.}~\bibnamefont {Marone}},\ and\ \bibinfo
		{author} {\bibfnamefont {M.}~\bibnamefont {Stampanoni}},\ }\bibfield  {title}
	{\bibinfo {title} {{X-ray} scattering tensor tomography with circular
			gratings},\ }\href@noop {} {\bibfield  {journal} {\bibinfo  {journal}
			{Applied Physics Letters}\ }\textbf {\bibinfo {volume} {116}},\ \bibinfo
		{pages} {134102} (\bibinfo {year} {2020})}\BibitemShut {NoStop}%
	\bibitem [{\citenamefont {Kim}\ \emph {et~al.}(2021)\citenamefont {Kim},
		\citenamefont {Kagias}, \citenamefont {Marone}, \citenamefont {Shi},\ and\
		\citenamefont {Stampanoni}}]{Kim2021}%
	\BibitemOpen
	\bibfield  {author} {\bibinfo {author} {\bibfnamefont {J.}~\bibnamefont
			{Kim}}, \bibinfo {author} {\bibfnamefont {M.}~\bibnamefont {Kagias}},
		\bibinfo {author} {\bibfnamefont {F.}~\bibnamefont {Marone}}, \bibinfo
		{author} {\bibfnamefont {Z.}~\bibnamefont {Shi}},\ and\ \bibinfo {author}
		{\bibfnamefont {M.}~\bibnamefont {Stampanoni}},\ }\bibfield  {title}
	{\bibinfo {title} {Fast acquisition protocol for {X-ray} scattering tensor
			tomography},\ }\href@noop {} {\bibfield  {journal} {\bibinfo  {journal}
			{Scientific reports}\ }\textbf {\bibinfo {volume} {11}},\ \bibinfo {pages}
		{23046} (\bibinfo {year} {2021})}\BibitemShut {NoStop}%
	\bibitem [{\citenamefont {Kim}\ \emph {et~al.}(2022{\natexlab{b}})\citenamefont
		{Kim}, \citenamefont {Pelt}, \citenamefont {Kagias}, \citenamefont
		{Stampanoni}, \citenamefont {Batenburg},\ and\ \citenamefont
		{Marone}}]{Kim2022}%
	\BibitemOpen
	\bibfield  {author} {\bibinfo {author} {\bibfnamefont {J.}~\bibnamefont
			{Kim}}, \bibinfo {author} {\bibfnamefont {D.~M.}\ \bibnamefont {Pelt}},
		\bibinfo {author} {\bibfnamefont {M.}~\bibnamefont {Kagias}}, \bibinfo
		{author} {\bibfnamefont {M.}~\bibnamefont {Stampanoni}}, \bibinfo {author}
		{\bibfnamefont {K.~J.}\ \bibnamefont {Batenburg}},\ and\ \bibinfo {author}
		{\bibfnamefont {F.}~\bibnamefont {Marone}},\ }\bibfield  {title} {\bibinfo
		{title} {{Tomographic Reconstruction of the Small-Angle X-Ray Scattering
				Tensor with Filtered Back Projection}},\ }\href@noop {} {\bibfield  {journal}
		{\bibinfo  {journal} {Physical Review Applied}\ }\textbf {\bibinfo {volume}
			{18}},\ \bibinfo {pages} {014043} (\bibinfo {year}
		{2022}{\natexlab{b}})}\BibitemShut {NoStop}%
	\bibitem [{\citenamefont {Felsner}\ \emph {et~al.}(2019)\citenamefont
		{Felsner}, \citenamefont {Hu}, \citenamefont {Maier}, \citenamefont {Bopp},
		\citenamefont {Ludwig}, \citenamefont {Anton},\ and\ \citenamefont
		{Riess}}]{Felsner2019}%
	\BibitemOpen
	\bibfield  {author} {\bibinfo {author} {\bibfnamefont {L.}~\bibnamefont
			{Felsner}}, \bibinfo {author} {\bibfnamefont {S.}~\bibnamefont {Hu}},
		\bibinfo {author} {\bibfnamefont {A.}~\bibnamefont {Maier}}, \bibinfo
		{author} {\bibfnamefont {J.}~\bibnamefont {Bopp}}, \bibinfo {author}
		{\bibfnamefont {V.}~\bibnamefont {Ludwig}}, \bibinfo {author} {\bibfnamefont
			{G.}~\bibnamefont {Anton}},\ and\ \bibinfo {author} {\bibfnamefont
			{C.}~\bibnamefont {Riess}},\ }\bibfield  {title} {\bibinfo {title} {{A 3-D
				Projection Model for X-ray Dark-field Imaging}},\ }\href@noop {} {\bibfield
		{journal} {\bibinfo  {journal} {Scientific Reports}\ }\textbf {\bibinfo
			{volume} {9}},\ \bibinfo {pages} {9216} (\bibinfo {year} {2019})}\BibitemShut
	{NoStop}%
	\bibitem [{\citenamefont {Graetz}(2021)}]{Graetz2021}%
	\BibitemOpen
	\bibfield  {author} {\bibinfo {author} {\bibfnamefont {J.}~\bibnamefont
			{Graetz}},\ }\bibfield  {title} {\bibinfo {title} {Simulation study towards
			quantitative x-ray and neutron tensor tomography regarding the validity of
			linear approximations of dark-field anisotropy},\ }\href@noop {} {\bibfield
		{journal} {\bibinfo  {journal} {Scientific Reports}\ }\textbf {\bibinfo
			{volume} {11}},\ \bibinfo {pages} {18477} (\bibinfo {year}
		{2021})}\BibitemShut {NoStop}%
	\bibitem [{\citenamefont {Aronson}\ and\ \citenamefont
		{Corngold}(1999)}]{Aronson1999}%
	\BibitemOpen
	\bibfield  {author} {\bibinfo {author} {\bibfnamefont {R.}~\bibnamefont
			{Aronson}}\ and\ \bibinfo {author} {\bibfnamefont {N.}~\bibnamefont
			{Corngold}},\ }\bibfield  {title} {\bibinfo {title} {Photon diffusion
			coefficient in an absorbing medium},\ }\href@noop {} {\bibfield  {journal}
		{\bibinfo  {journal} {J Opt Soc Am A Opt Image Sci Vis}\ }\textbf {\bibinfo
			{volume} {16}},\ \bibinfo {pages} {1066} (\bibinfo {year}
		{1999})}\BibitemShut {NoStop}%
	\bibitem [{\citenamefont {Paganin}\ and\ \citenamefont
		{Morgan}(2019)}]{Paganin2019}%
	\BibitemOpen
	\bibfield  {author} {\bibinfo {author} {\bibfnamefont {D.~M.}\ \bibnamefont
			{Paganin}}\ and\ \bibinfo {author} {\bibfnamefont {K.~S.}\ \bibnamefont
			{Morgan}},\ }\bibfield  {title} {\bibinfo {title} {X-ray fokker–planck
			equation for paraxial imaging},\ }\href@noop {} {\bibfield  {journal}
		{\bibinfo  {journal} {Scientific Reports}\ }\textbf {\bibinfo {volume} {1}},\
		\bibinfo {pages} {17537} (\bibinfo {year} {2019})}\BibitemShut {NoStop}%
	\bibitem [{sup()}]{supplementary}%
	\BibitemOpen
	\href@noop {} {}\bibinfo {note} {See Supplemental Material at [{URL} will be
		inserted by publisher] for the derivation of the adjoint operator, and the
		derivation of projection operator}\BibitemShut {NoStop}%
	\bibitem [{\citenamefont {Basser}\ and\ \citenamefont
		{Pierpaoli}(1996)}]{Basser1996}%
	\BibitemOpen
	\bibfield  {author} {\bibinfo {author} {\bibfnamefont {P.~J.}\ \bibnamefont
			{Basser}}\ and\ \bibinfo {author} {\bibfnamefont {C.}~\bibnamefont
			{Pierpaoli}},\ }\bibfield  {title} {\bibinfo {title} {Microstructural and
			physiological features of tissues elucidated by quantitative-diffusion-tensor
			{MRI}},\ }\href@noop {} {\bibfield  {journal} {\bibinfo  {journal} {Journal
				of Magnetic Resonance B}\ }\textbf {\bibinfo {volume} {111}},\ \bibinfo
		{pages} {209} (\bibinfo {year} {1996})}\BibitemShut {NoStop}%
	\bibitem [{\citenamefont {Shi}\ \emph {et~al.}(2022)\citenamefont {Shi},
		\citenamefont {Josell}, \citenamefont {Jefimovs}, \citenamefont {Romano},
		\citenamefont {Moffat}, \citenamefont {Stampanoni},\ and\ \citenamefont
		{Schlepütz}}]{Shi2022}%
	\BibitemOpen
	\bibfield  {author} {\bibinfo {author} {\bibfnamefont {Z.}~\bibnamefont
			{Shi}}, \bibinfo {author} {\bibfnamefont {D.}~\bibnamefont {Josell}},
		\bibinfo {author} {\bibfnamefont {K.}~\bibnamefont {Jefimovs}}, \bibinfo
		{author} {\bibfnamefont {L.}~\bibnamefont {Romano}}, \bibinfo {author}
		{\bibfnamefont {T.~P.}\ \bibnamefont {Moffat}}, \bibinfo {author}
		{\bibfnamefont {M.}~\bibnamefont {Stampanoni}},\ and\ \bibinfo {author}
		{\bibfnamefont {C.~M.}\ \bibnamefont {Schlepütz}},\ }\bibfield  {title}
	{\bibinfo {title} {Fabrication of a fractal pattern device for focus
			characterizations of {X-ray} imaging systems by {Si} deep reactive ion
			etching and bottom-up {Au} electroplating},\ }\href@noop {} {\bibfield
		{journal} {\bibinfo  {journal} {Applied Optics}\ }\textbf {\bibinfo {volume}
			{61}},\ \bibinfo {pages} {3850} (\bibinfo {year} {2022})}\BibitemShut
	{NoStop}%
	\bibitem [{\citenamefont {Kagias}\ \emph {et~al.}(2016)\citenamefont {Kagias},
		\citenamefont {Wang}, \citenamefont {Villanueva-Perez}, \citenamefont
		{Jefimovs},\ and\ \citenamefont {Stampanoni}}]{Kagias2016}%
	\BibitemOpen
	\bibfield  {author} {\bibinfo {author} {\bibfnamefont {M.}~\bibnamefont
			{Kagias}}, \bibinfo {author} {\bibfnamefont {Z.}~\bibnamefont {Wang}},
		\bibinfo {author} {\bibfnamefont {P.}~\bibnamefont {Villanueva-Perez}},
		\bibinfo {author} {\bibfnamefont {K.}~\bibnamefont {Jefimovs}},\ and\
		\bibinfo {author} {\bibfnamefont {M.}~\bibnamefont {Stampanoni}},\ }\bibfield
	{title} {\bibinfo {title} {{2D-Omnidirectional Hard-X-Ray Scattering
				Sensitivity in a Single Shot}},\ }\href@noop {} {\bibfield  {journal}
		{\bibinfo  {journal} {Physical Review Letters}\ }\textbf {\bibinfo {volume}
			{116}},\ \bibinfo {pages} {093902} (\bibinfo {year} {2016})}\BibitemShut
	{NoStop}%
	\bibitem [{\citenamefont {Kagias}\ \emph {et~al.}(2019)\citenamefont {Kagias},
		\citenamefont {Wang}, \citenamefont {Birkbak}, \citenamefont {Lauridsen},
		\citenamefont {Abis}, \citenamefont {Lovric}, \citenamefont {Jefimovs},\ and\
		\citenamefont {Stampanoni}}]{Kagias2019}%
	\BibitemOpen
	\bibfield  {author} {\bibinfo {author} {\bibfnamefont {M.}~\bibnamefont
			{Kagias}}, \bibinfo {author} {\bibfnamefont {Z.}~\bibnamefont {Wang}},
		\bibinfo {author} {\bibfnamefont {M.~E.}\ \bibnamefont {Birkbak}}, \bibinfo
		{author} {\bibfnamefont {E.}~\bibnamefont {Lauridsen}}, \bibinfo {author}
		{\bibfnamefont {M.}~\bibnamefont {Abis}}, \bibinfo {author} {\bibfnamefont
			{G.}~\bibnamefont {Lovric}}, \bibinfo {author} {\bibfnamefont
			{K.}~\bibnamefont {Jefimovs}},\ and\ \bibinfo {author} {\bibfnamefont
			{M.}~\bibnamefont {Stampanoni}},\ }\bibfield  {title} {\bibinfo {title}
		{Diffractive small angle {X-ray} scattering imaging for anisotropic
			structures},\ }\href@noop {} {\bibfield  {journal} {\bibinfo  {journal}
			{Nature Communications}\ }\textbf {\bibinfo {volume} {10}},\ \bibinfo {pages}
		{5130} (\bibinfo {year} {2019})}\BibitemShut {NoStop}%
	\bibitem [{\citenamefont {Mokso}\ \emph {et~al.}(2017)\citenamefont {Mokso},
		\citenamefont {Schlepütz}, \citenamefont {Theidel}, \citenamefont {Billich},
		\citenamefont {Schmid}, \citenamefont {Celcer}, \citenamefont {Mikuljan},
		\citenamefont {Sala}, \citenamefont {Marone}, \citenamefont {Schlumpf},\ and\
		\citenamefont {Stampanoni}}]{Mokso2017}%
	\BibitemOpen
	\bibfield  {author} {\bibinfo {author} {\bibfnamefont {R.}~\bibnamefont
			{Mokso}}, \bibinfo {author} {\bibfnamefont {C.~M.}\ \bibnamefont
			{Schlepütz}}, \bibinfo {author} {\bibfnamefont {G.}~\bibnamefont {Theidel}},
		\bibinfo {author} {\bibfnamefont {H.}~\bibnamefont {Billich}}, \bibinfo
		{author} {\bibfnamefont {E.}~\bibnamefont {Schmid}}, \bibinfo {author}
		{\bibfnamefont {T.}~\bibnamefont {Celcer}}, \bibinfo {author} {\bibfnamefont
			{G.}~\bibnamefont {Mikuljan}}, \bibinfo {author} {\bibfnamefont
			{L.}~\bibnamefont {Sala}}, \bibinfo {author} {\bibfnamefont {F.}~\bibnamefont
			{Marone}}, \bibinfo {author} {\bibfnamefont {N.}~\bibnamefont {Schlumpf}},\
		and\ \bibinfo {author} {\bibfnamefont {M.}~\bibnamefont {Stampanoni}},\
	}\bibfield  {title} {\bibinfo {title} {{GigaFRoST}: The gigabit fast readout
			system for tomography},\ }\href@noop {} {\bibfield  {journal} {\bibinfo
			{journal} {Journal of Synchrotron Radiation}\ }\textbf {\bibinfo {volume}
			{24}},\ \bibinfo {pages} {1250} (\bibinfo {year} {2017})}\BibitemShut
	{NoStop}%
	\bibitem [{\citenamefont {De~Marco}\ \emph {et~al.}(2023)\citenamefont
		{De~Marco}, \citenamefont {Savatović}, \citenamefont {Smith}, \citenamefont
		{Di~Trapani}, \citenamefont {Margini}, \citenamefont {Lautizi},\ and\
		\citenamefont {Thibault}}]{DeMarco2023}%
	\BibitemOpen
	\bibfield  {author} {\bibinfo {author} {\bibfnamefont {F.}~\bibnamefont
			{De~Marco}}, \bibinfo {author} {\bibfnamefont {S.}~\bibnamefont
			{Savatović}}, \bibinfo {author} {\bibfnamefont {R.}~\bibnamefont {Smith}},
		\bibinfo {author} {\bibfnamefont {V.}~\bibnamefont {Di~Trapani}}, \bibinfo
		{author} {\bibfnamefont {M.}~\bibnamefont {Margini}}, \bibinfo {author}
		{\bibfnamefont {G.}~\bibnamefont {Lautizi}},\ and\ \bibinfo {author}
		{\bibfnamefont {P.}~\bibnamefont {Thibault}},\ }\bibfield  {title} {\bibinfo
		{title} {High-speed processing of {X-ray} wavefront marking data with the
			{Unified Modulated Pattern Analysis (UMPA)} model},\ }\href@noop {}
	{\bibfield  {journal} {\bibinfo  {journal} {Optics Express}\ }\textbf
		{\bibinfo {volume} {31}},\ \bibinfo {pages} {635} (\bibinfo {year}
		{2023})}\BibitemShut {NoStop}%
\end{thebibliography}
\providecommand{\noopsort}[1]{}\providecommand{\singleletter}[1]{#1}%

\clearpage
\onecolumngrid
\appendix

\preprint{APS/123-QED}

\title{A universal reconstruction method for X-ray scattering tensor tomography based on wavefront modulation
\\ -- \textit{Supplemental material} --}
\author{Ginevra Lautizi\textsuperscript{1,2}}
\email{ginevra.lautizi@phd.units.it}
\author{Alain Studer\textsuperscript{3}}
\author{Marie-Christine Zdora\textsuperscript{4,5}}
\author{Fabio De Marco\textsuperscript{1,2}}
\author{Jisoo Kim\textsuperscript{6}}
\author{Vittorio  Di Trapani\textsuperscript{1,2}}
\author{Federica Marone\textsuperscript{4}}
\author{Pierre Thibault\textsuperscript{1,2}}
\author{Marco Stampanoni\textsuperscript{4,5}}
\affiliation{\textsuperscript{1} Department of Physics, University of Trieste, Trieste, Italy }
\affiliation{\textsuperscript{2} Elettra-Sincrotrone Trieste, Basovizza, Italy}
\affiliation{\textsuperscript{3} Data Processing Development and Consulting Group, Paul Scherrer Institut, Villigen 5232, Switzerland}
\affiliation{\textsuperscript{4} Photon Science Division, Paul Scherrer Institut, Villigen 5232, Switzerland}
\affiliation{\textsuperscript{5} Institute for Biomedical Engineering, ETH Zürich, Zürich 8092, Switzerland}
\affiliation{\textsuperscript{6}Advanced Instrumentation Institute, Korea Research Institute of Standards and Science, Korea}

\maketitle

\end{comment}

\section{The adjoint operator}
As stated in the manuscript, the adjoint operator 
$A^*: L^2(\mathbb{R}^{4}, \mathbb{R}^{3 \times 3}) \rightarrow
L^2(\mathbb{R}^{3}, \mathbb{R}^{3 \times 3})$
equals

\begin{equation}
A^*[S](\bm{r}) =
 \int_0^{2\pi} \int_0^{\pi/2}
 (R_{\beta}^x R_{\alpha}^z)^{-1} S(P_B R_{\beta}^x R_{\alpha}^z \bm{r}, \alpha, \beta)  R_{\beta}^x R_{\alpha}^z \, \cos \beta d \beta d \alpha.
\end{equation}
It can be proved as follows.
The inner product in both spaces is defined as
$\langle X, Y \rangle = \int tr(XY^T)$. 
Hence for a sinogram $S$ and a tensor field $F$, the inner product 
$\langle S, A [F] \rangle$ equals

\begin{equation}
\int_{\mathbb{R}^2} \int_0^{2\pi} \int_0^{\pi/2}
tr \big\{S(P_B \bm{r}, \alpha, \beta) \int_{\mathbb{R}}
[P_B D(R_{\beta}^x)D(R_{\alpha}^z)[F] (\bm{r}) P_B]^T dy\big\} 
\,  d \beta d \alpha dz dx.
\end{equation}
Using the linearity of the trace and the transformation rule for tensor fields

\begin{equation}
\int_{\mathbb{R}^3} \int_0^{2\pi} \int_0^{\pi/2}
tr \big\{S(P_B \bm{r}, \alpha, \beta) P_B R_{\beta}R_{\alpha}F^T(R_{\alpha}^T R_{\beta}^T\bm{r}) R_{\alpha}^T  R_{\beta}^T P_B \big\} \, d \beta d \alpha d \bm{r}.
\end{equation}
The trace is cyclic, such that

\begin{equation}
\int_{\mathbb{R}^3} \int_0^{2\pi} \int_0^{\pi/2}
tr \big\{ R_{\alpha}^T R_{\beta}^T P_B T(P_B \bm{r}, \alpha, \beta) P_B R_{\beta} R_{\alpha}F^T(R_{\alpha}^T R_{\beta}^T\bm{r}) \big\} \, d \beta d \alpha d \bm{r},
\end{equation}

shifting the rotations to the sinogram and again skipping the projector, since the sinogram is already projected.
\begin{equation}
\int_{\mathbb{R}^3} \int_0^{2\pi} \int_0^{\pi/2}
tr \big\{ R_{\alpha}^T R_{\beta}^T S(P_B R_{\beta} R_{\alpha} \bm{r}, \alpha, \beta) R_{\beta} R_{\alpha}F^T(\bm{r}) \big\} \, d \beta d \alpha d \bm{r}.
\end{equation}

Due to linearity of the trace, we can rearrange this as 
\begin{equation}
\int_{\mathbb{R}^3} 
tr \Big\{ \Big(\int_0^{2\pi} \int_0^{\pi/2} R_{\alpha}^T  R_{\beta}^T S(P_B R_{\beta} R_{\alpha} \bm{r}, \alpha, \beta)  R_{\beta}R_{\alpha} d \alpha d \beta \Big) F^T(\bm{r}) \Big\} \, d \bm{r}.
\end{equation}

This must equal by definition

\begin{equation}
\langle  A^*[S], F \rangle =
\int_{\mathbb{R}^3} 
tr \Big\{  A^* [S](\bm{r})  F^T(\bm{r}) \Big\} \, d \bm{r},
\end{equation}
from which we see that

\begin{equation}
 A^*[S](\bm{r}) =
 \int_0^{2\pi} \int_0^{\pi/2}
 (R_{\beta}^x R_{\alpha}^z)^{-1} S(P_B R_{\beta}^x R_{\alpha}^z \bm{r}, \alpha, \beta)  R_{\beta}^x R_{\alpha}^z 
 \, d \beta d \alpha.
\end{equation}
Instead of choosing the simple integral measure $d \beta$,
we can as well chose the measure $\cos \beta d \beta$.
\newline
Indeed, for $\beta = \pi/2$, the sample is rotated around the beam axis, making tomography redundant. One single $\alpha$ value, e.g $\alpha=0$, is sufficient to capture all information available. This fact is covered with the $\cos \beta$ factor, by which the number of $\alpha$ rotations can be decreased while increasing $\beta$.
This completes the proof.

\section{Derivation of Projection Operator $P_y$}
We will here prove that the detector projects the full 3D tensor $\Sigma$ to the 2D version $P_y \Sigma :=P_B \Sigma P_B^T$.
Starting from Eq.~\ref{2d_real} of the main manuscript and considering only one scattering layer:
\begin{equation}
I_{scatter} = I_{0} *_{3D} MVG(\Sigma)
\end{equation}
where MVG stands for Multi Variate Gaussian and $I_{scatter}(\bm{r})$ can be seen as the 3D intensity distribution arriving at the detector, after having exited the thin mono-layer sample.

The intensity measured at the detector reads

\begin{equation}
I_{det}(x,z) =
\int I_{scatter}(x,y,z) dy =
\int \int I_{0}((x,y,z) - \bm{r'}) dy \,
MVG(\bm{r'}; \Sigma)  \, d \bm{r'},
\end{equation}
where we have explicitly written the convolution integral. Since the support of the blurring probability density function is much smaller than the the size of the unit cell support, we can approximate $I_{0}(x -x',-y',z-z')$ as a constant with respect to $y'$ (evaluated at zero) such that

\begin{equation}
I_{det}(x,z) \simeq
\int \int I_{0}(x - x', y , z - z') dy \,
 \Big\{ \int MVG(\bm{r'}; \Sigma)dy'   \Big\}
 \, dx' dz'.
\end{equation}

Defining the projection
operator as $P_y \Sigma := P_B \Sigma P_B^T$, it holds that a MVG probability density function is still a MVG, with dimension reduced accordingly:

\begin{equation}
\int MVG(x,y,z; \Sigma)dy =
MVG(x, P_y \Sigma, z).
\end{equation}
From this we see that

\begin{equation}
I_{det} = I_0 *_{2D} MVG(P_y \Sigma),
\end{equation}
where $I_0(x,z) = \int I_{0}(x,y,z)dy$. 
This completes the proof.


\end{document}